\documentclass[cernpreprint, REPORT=false, atlasdraft=false, UKenglish, texmf, orcidlogo]{atlasdoc}
\usepackage[subfigure]{atlaspackage}
\usepackage{atlasbiblatex}
\usepackage{atlascontribute}

\usepackage[BSM=true,jetetmiss=true,hepprocess=true]{atlasphysics}
\usepackage{ANA-SUSY-2023-10-PAPER-defs}
\graphicspath{{logos/}{figures/}}

\usepackage{ANA-SUSY-2023-10-PAPER-defs}

\AtlasTitle{The quest to discover supersymmetry at the ATLAS experiment}

\AtlasAbstract{%
The search for supersymmetry with the ATLAS experiment at the CERN Large Hadron Collider intensified after the discovery of the Higgs boson in 2012. The search programme expanded in both breadth and depth, profiting from the increased integrated luminosity and higher centre-of-mass energy for the collision data collected between 2015 and 2018, and gaining new sensitivity to unexplored areas of supersymmetry parameter space through the use of novel experimental signatures and innovative analysis techniques. This report summarises the supersymmetry searches at ATLAS using up to 140 fb$^{-1}$ of $pp$ collisions at $\sqrt{s} = 13$ TeV, including the limits set on the production of gluinos, squarks, and electroweakinos for scenarios with or without R-parity conservation, and including models where some of the supersymmetric particles are long-lived.
}

\author{The ATLAS Collaboration}

\AtlasRefCode{SUSY-2023-10}
\AtlasJournal{Physics Reports}

\PreprintIdNumber{CERN-EP-2024-056}

\AtlasRefCode{SUSY-2023-10}

\AtlasNote{SUSY-2023-10}

\AtlasJournalRef{Phys. Rep. 1116 (2025) 261-300}
\AtlasDOI{https://doi.org/10.1016/j.physrep.2024.09.010}

\hypersetup{pdftitle={ATLAS document},pdfauthor={The ATLAS Collaboration}}

\begin{document}

\maketitle

\tableofcontents

\section{Introduction}
%

%

One of the most significant contributions of the CERN Large Hadron Collider (LHC)~\cite{Evans:2008zzb} to high-energy physics comes through the particles that the ATLAS~\cite{PERF-2007-01} and CMS~\cite{CMS-CMS-00-001} collaborations have {\em not} found. Both collaborations have pursued an unprecedented programme of searches for phenomena not predicted by the Standard Model (SM). The wide variety of signatures explored, and the richness of the models considered, has had a powerful influence on community's paradigms of physics beyond the SM.

Among these paradigms, supersymmetry (SUSY)~\cite{Golfand:1971iw,Volkov:1973ix,Wess:1974tw,Wess:1974jb,Ferrara:1974pu,Salam:1974ig} is one of the most closely examined. The approach of imposing symmetries on Lagrangians led to the construction of electroweak theory, the unification of the weak and electromagnetic interaction and, eventually, the development of the SM. The phenomenology of SUSY stems from requiring the Lagrangian to be invariant under an operator that maps fermionic fields into bosonic ones, and vice versa. It was found that only additional space-time symmetry could be added to the Poincar\'e group~\cite{Volkov:1973ix}. To impose this symmetry, one needs to add new {\em superpartners} ofor each SM particle. The much richer particle content, and some of the free parameters that one needs to add to make SUSY a broken symmetry (for example, it is known that there is no scalar partner of the electron with its same mass), makes SUSY an ideal framework to accommodate many of the shortcomings of the SM. The quantum corrections to the Higgs boson mass coming from the fermions are counterbalanced by those coming from their superpartners, stabilising the mass to a value near the electroweak scale in a more natural way.
On top of that, the modified particle content changes the evolution of the running gauge couplings of the SM, potentially allowing them to converge to a common value, a necessary condition for unification of the electroweak and strong interactions. The additional particles come with new mixing parameters, possibly introducing  additional sources of CP violation and thus paving the way for an explanation of the cosmological matter--antimatter asymmetry. SUSY even offers the attractive possibility of generating electroweak symmetry breaking dynamically. Finally, there may be room for one or more dark-matter candidates among the particles introduced in the model.

Because of its elegance and flexibility, SUSY has received significant theoretical and experimental attention in the past decades. The generic prediction of the existence of SUSY particles at the \TeV scale created widespread expectations for experimental results from the LHC collaborations in the early 2000s. The ATLAS Collaboration rose to these expectations by developing a thorough, robust and extensive experimental effort to search for SUSY in data from \RunOne (2010 to 20013) , at $\sqrt{s} = 7$~and~$8\,\TeV$, and \RunTwo (2015-2018), at $\sqrt{s} = 13\,\TeV$. The aim of this paper is to review and summarise this effort using data collected during \RunTwo of the LHC. Section~\ref{sec:models} gives a quick overview of the main theoretical and phenomenological aspects of supersymmetry, introducing the approach ATLAS used to explore  the vast SUSY parameter space.
Section~\ref{sec:atlasdet} provides a brief description of the ATLAS dectector before
Section~\ref{sec:strategy} discusses generic aspects related to the design of SUSY searches, the background estimation, and the statistical approach used to interpret the search results and set constraints on sparticle masses, couplings, and production cross-sections.

The searches described in this paper covered a large variety of experimental signatures arising from models featuring a stable weakly interacting particle (the R-parity-conserving, or RPC, models) and from models without such a potential dark-matter candidate (R-parity-violating, or RPV, models). Both types of models are described in Section~\ref{sec:models}. Sections~\ref{sec:Strong} and \ref{sec:Weak} review the analyses targeting the production of SUSY particles via strong or electroweak interactions, respectively, when using RPC SUSY models: generically speaking, the common feature of these searches is the presence of missing transverse momentum $\ptmiss$ (with magnitude $\met$) in signal events, arising from the presence of a dark-matter candidate, which goes undetected. Section~\ref{sec:RPV} discusses RPV SUSY searches, where an $\met$ signature is typically absent. Experimental signatures involving long-lived particles arise quite naturally in some regions of the parameter space: Section~\ref{sec:LLP} focuses on the effort to target such signatures with unconventional experimental techniques.

Each of the searches discussed in these sections makes use of simplified models (discussed in Section~\ref{sec:models}) to optimise the  analysis and interpret the results. While the simplified-model approach gives a direct connection to the signal topologies, it has the drawback of yielding mass limits that rarely generalise to more complex/complete SUSY models. Section~\ref{sec:pmssm} focuses on reinterpreting the analyses designed using simplified models in the more general phenomenological Minimal
Supersymmetric Standard Model (pMSSM), with the aim of giving an overall picture of the actual impact of ATLAS on the supersymmetric landscape.

The authors hope that this paper will serve as a useful introduction to the field for those wanting to take an active role in research activities connected to SUSY. At the same time, it may be a handy collection of information and references for the HEP community, either for easy access to ATLAS research on the subject or simply as an overview of the searches performed.


%
\section{Supersymmetric models}
\label{sec:models}
%

%

To make the SM supersymmetric,\footnote{Non-minimal extensions of the SM can be obtained by first extending the SM and then completing it by introducing the supersymmetric partners of the SM fields, or by considering multiple versions of the SUSY operators applied to the SM fields.} one has first to postulate the existence of an extended Higgs field sector: two complex doublets provide the minimum number of fields that can give mass to the up-type and down-type particles without spoiling gauge invariance. The next step is to introduce the supersymmetric partners of each SM field: for each fermion $f$ (characterised by a left-handed and a right-handed component, $f_{\mathrm{L}}$ and $f_{\mathrm{R}}$, respectively), one introduces two supersymmetric scalar fields $\tilde{f}_{\mathrm{L}}$ and $\tilde{f}_{\mathrm{R}}$. Although the subscripts $\mathrm{L}$ and $\mathrm{R}$ have nothing to do with chirality at this point, one often refers to these scalar fields as the {\em left-} and {\em right-chiral sfermions}.\footnote{The supersymmetric scalars (partners of SM fermions) take the name of the SM partner with an \enquote{s} prepended. For example, the electron has two partner {\em selectrons}. The supersymmetric fermions (partners of SM bosons) take the name of their SM partner with an \enquote{ino} suffix. For example, the supersymmetric partner of the gluon is the {\em gluino}.} These two fields mix to yield the mass eigenstates $\tilde{f}_1$ and $\tilde{f}_2$, with $\mass{\tilde{f}_1} <  \mass{\tilde{f}_2}$ by convention. For the electroweak boson fields, one starts from 16 spin degrees of freedom before electroweak symmetry breaking (two each for the $B$ and $\mathbf{W}$ massless spin-1 fields, eight corresponding to the two complex Higgs doublets). This requires the introduction of eight supersymmetric fermionic partners: one {\em bino}, three {\em winos} and four {\em higgsinos}.  The electroweak symmetry breaking results in the usual three massive vector bosons $W^{\pm}$ and $Z$, and the massless $\gamma$ (11 degrees of freedom), plus five spin-0 bosons ($h$, $H$, $A$, $H^{\pm}$; often, $h$ is used to denote the SM-like 125 GeV Higgs boson). The bino, winos and higgsinos mix, yielding eight fermionic mass eigenstates (the neutralinos $\tilde{\chi}_i^0$, with $i = [1,4]$, and the charginos $\tilde{\chi}_1^{\pm}$ and $\tilde{\chi}_2^{\pm}$, with the subscript increasing with increasing mass), corresponding to 16 degrees of freedom. Finally, the fermionic {\em gluino}, $\gluino$, is introduced as a superpartner of the gluon, and in models where local SUSY is broken by the superHiggs mechanism, the gravitino, $\gravino$, is the fermionic superpartner of the hypothetical graviton.

The particle content of the SM is therefore more than doubled by its supersymmetrisation. Supersymmetry requires that the couplings of the SUSY partners are identical to the corresponding ones in the SM (after accounting for the extended Higgs sector). If supersymmetry were an exact symmetry of nature, then the SUSY particle masses would be equal to those of the corresponding SM partners. The absence, so far, of any observed SUSY particle implies that SUSY must be a broken symmetry.
The SUSY breaking mechanism, although unknown, can modify the phenomenology and mass spectrum of the SUSY sector.
For example, conventional theories of gravity-mediated SUSY breaking, \enquote{supergravity} (SUGRA) theories, include massive gravitinos due to the superHiggs mechanism.
Gauge-mediated SUSY breaking (GMSB) theories usually have a nearly massless gravitino as the lightest supersymmetric particle (LSP), while anomaly-mediated SUSY breaking (AMSB) models usually feature long-lived $\chinoonepm$ decays to pure-wino $\ninoone$ LSPs.
The scale at which SUSY is broken is inherently tied to the breaking mechanism, and also impacts naturalness arguments that favour light top squarks, higgsinos, and moderately light gluinos to avoid excessive tuning of the weak scale.
The complexity of defining the mechanism and scale of SUSY breaking is usually avoided by incorporating effective {\em soft-SUSY-breaking} terms in the model. They include many parameters: 105, including supersymmetric particle masses and field phases that cannot be absorbed by a redefinition of the fields. The model sketched so far is called the Minimal Supersymmetric Standard Model (MSSM)~\cite{Baer_Tata_2006}.

Vast regions of the MSSM parameter space are not consistent with the observed particle physics phenomenology. Further assumptions motivated by experimental results are often made. To prevent the proton from decaying too quickly~\cite{Farrar:1978xj}, conservation of a multiplicative quantum number called R-parity is often assumed. The R-parity of a particle is a function of its baryon number $B$, the lepton number $L$, and the spin $S$, and is defined as
\begin{align*}
R= (-1)^{3(B-L)+2S},
\end{align*}
giving a value of $-1$ for SUSY particles and $+1$ for SM particles. Its conservation has important phenomenological consequences because an even number of SUSY particles must always appear at a Feynman vertex. This implies that SUSY particles must be produced in pairs, and that the LSP is stable. A necessary condition for the LSP to be a good dark-matter candidate is that it interacts with ordinary matter only through weak interactions, which implies that LSPs produced (either directly or from the decay of a SUSY particle) in $pp$ collisions do not interact with the detector, and $\ptmiss$ becomes a characteristic event signature.

While R-parity conservation gives rise to a compelling phenomenology, it is not dictated by any fundamental principle. R-parity-violating (RPV) couplings are allowed, as long as their magnitude is not incompatible with existing observations. Because of this constraint, RPV couplings are typically small enough to be neglected in the production of SUSY particles, but they can have important consequences for SUSY particle decays: in particular, the LSP can decay into SM particles. This eliminates, in general, the possibility of retaining a dark-matter candidate, and opens many options for the identity of the LSP, which can essentially be any SUSY particle. Since the LSP can decay, there is no expectation of having significant $\ptmiss$ in the final state. RPV SUSY delivers very different event topologies than RPC SUSY.

Other assumptions can be made in order to reduce the volume of parameter space to be explored in the MSSM. Relatively general and well-motivated assumptions~\cite{Djouadi:2002ze,Berger:2008cq,MSSMWorkingGroup:1998fiq} about the flavour and mass structures of the MSSM allow the number of free parameters to be reduced to 19 or 20 (depending on the nature of the LSP): this so-called phenomenological MSSM (pMSSM) was investigated extensively during \RunOne of the LHC by both ATLAS~\cite{SUSY-2014-08} and CMS~\cite{CMS-SUS-15-010}.

The strategy chosen by ATLAS to map its search programme to such a vast parameter space was to make use of simplified models~\cite{Alwall:2008ve,Alwall:2008ag,Alves:2011wf}. In a simplified model, one assumes that only a handful of SUSY particles are relevant for the phenomenology of $pp$ collisions. It is often assumed that a single type of SUSY particle is produced, with its decay involving only a few SUSY particles, or only the LSP. Designing an analysis for a specific simplified model means targeting a specific event topology. The limits obtained for simplified models hold whenever the assumed particle mass hierarchy and decays are realised in a fully developed SUSY model (either the MSSM or some more complex model). The main drawback of this approach is that it is often difficult to generalise a mass exclusion limit obtained for a SUSY particle\footnote{A danger of simplified models is that they may lead to neglect constrains arising from other sectors of the theory. For example, GMSB models lead to sets of viable simplified models with a gravitino LSP, but require stops with very large masses to yield $\mass{h}= 125 \ \GeV$.}. In a general SUSY model, many production processes and decay chains often compete with each other, and sometimes interference structures need to be taken into account when computing cross-sections. In short, the complexity of the actual model clashes with the clean and simple assumptions of the simplified models, so the simplified-model limits should always be presented along with caveats due to the assumptions made in designing the models, namely the (typically single) production process and decay mode.

\begin{figure}[t]
\begin{center}
\subfigure[]{
\includegraphics[width=0.28\textwidth]{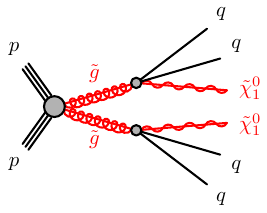}
\label{fig:feyn_diag_a}}
\subfigure[]{
\includegraphics[width=0.28\textwidth]{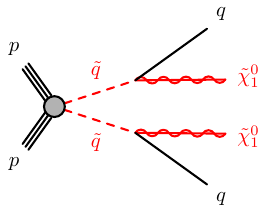}\label{fig:feyn_diag_b}
}
\subfigure[]{
\includegraphics[width=0.28\textwidth]{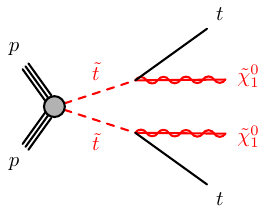}\label{fig:feyn_diag_c}
}
\end{center}
\caption{Simplified models of (a) gluino pair production followed by $\gluino\rightarrow  q\bar{q}\ninoone$; (b)  squark pair production followed by $\squark \rightarrow q \ninoone$; and (c) top squark (stop) pair production followed by $\stop \rightarrow t^{(*)} \ninoone$. No distinction between particle and antiparticle is done in these diagrams.}
\end{figure}

The production cross-sections for strongly interacting SUSY particles hardly depend on any SUSY model parameter beside the sparticle masses. This is a consequence of the SU(3) gauge interaction driving the process. The Feynman diagrams involved in strong production are the equivalent of those for production of gluons and quarks in the SM. At the next-to-leading order in the strong coupling constant \alphas  and beyond, interference between diagrams containing squarks and gluinos makes the production cross-section for each of the relevant processes ($\gluino\gluino$, $\gluino\squark$, $\squark\squark$) dependent on the specific mass spectrum of the model. For example, destructive interference between $t$-channel exchange of $\gluino$ and $\squark$ makes the $\gluino\gluino$ production cross-section dependent on the $\squark$ masses. However, if the other SUSY particles are assumed not to take part (not even virtually) in the production process, then the production cross-sections depend only on the sparticle mass. These are the assumptions typically made in simplified models of gluino or squark pair production, shown in Figures~\ref{fig:feyn_diag_a} and~\ref{fig:feyn_diag_b} respectively, and in Figure~\ref{fig:feyn_diag_c} if the squark is a top squark (or \enquote{stop}).

\begin{figure}[t]
\centering
\includegraphics[width=0.6\textwidth]{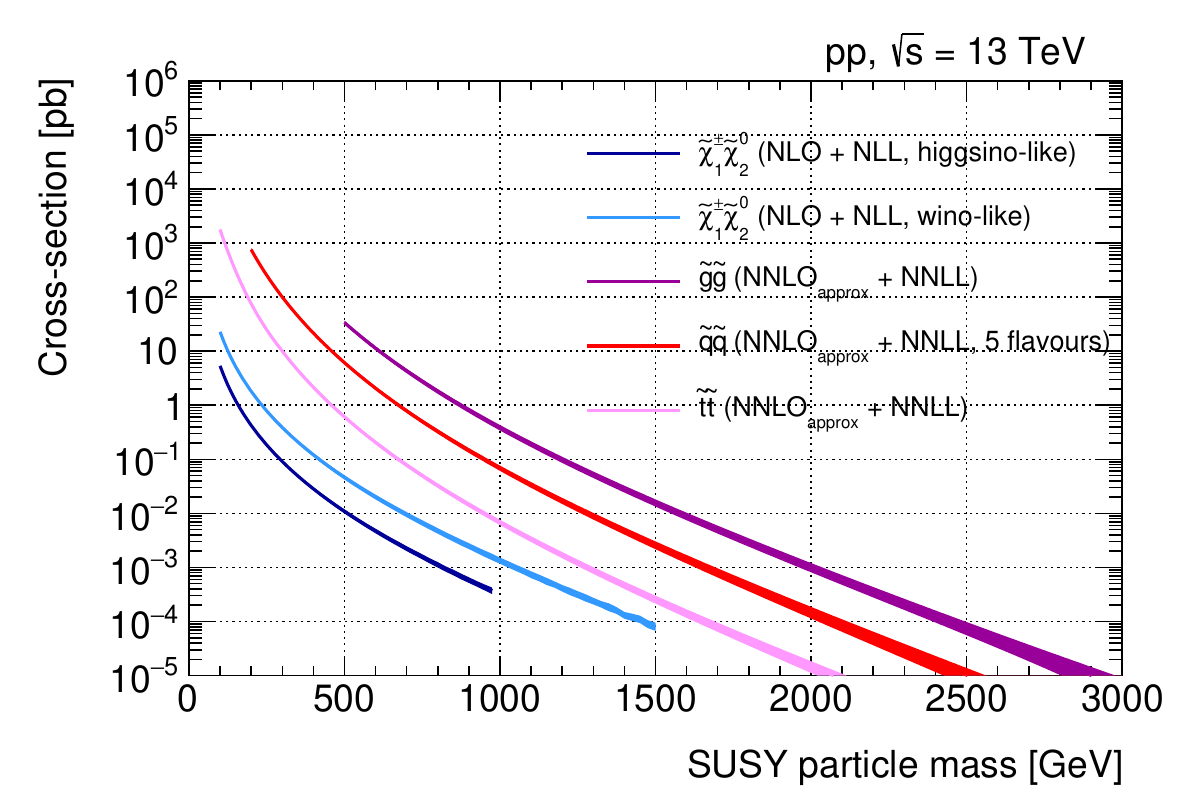}
\caption{Direct pair-production cross-sections for a few processes mediated by the strong or electroweak processes. The width of the line represents the theoretical uncertainty. The values of the cross-sections and uncertainties are those agreed upon by the LHC SUSY Cross Section Working Group~\cite{susyXsec}.
}
\label{fig:xsec-plot}
\end{figure}

The production cross-sections for squarks and gluinos with simplified-model assumptions are shown in Figure~\ref{fig:xsec-plot}. The cross-section shown for squark production assumes that the 10 squarks corresponding to the $u,d,c,s$, and $b$ flavours (and two chirality states) are all degenerate in mass. The cross-section scales linearly with the number of squark types that can be produced.

For example, the production cross-section for a pair of gluinos with $\mass{\gluino} = 2\,\TeV$ is about 1\,fb, assuming that the mass of all other SUSY particles is large enough for them to be irrelevant in the cross-section evaluation. Consequently, between 100 and 200 gluinos with $m_{\gluino} = 2\,\TeV$ would have been produced at the ATLAS interaction point during \RunTwo. The production of SUSY particles with such a large mass typically results in a quite distinctive event topology, so that the event selection can achieve sufficient background rejection while retaining relatively large signal acceptance and efficiency. It is therefore not surprising that  ATLAS is typically sensitive to gluinos in the mass range up to $m_{\gluino} = 2\,\TeV$ or somewhat higher, unless specific mass hierarchies are realised such that event selections with very small acceptance and/or efficiency need to be deployed. Following the same logic, one sees that the sensitivity of ATLAS to tenfold-degenerate squark pair production reaches masses approximately 0.5\,TeV lower than those corresponding to gluino pair production.

\begin{figure}[t]
\begin{center}
\subfigure[]{
\includegraphics[width=0.28\textwidth]{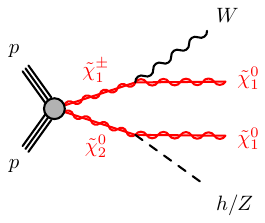}\label{fig:feyn_diag_d}
}
\subfigure[]{
\includegraphics[width=0.28\textwidth]{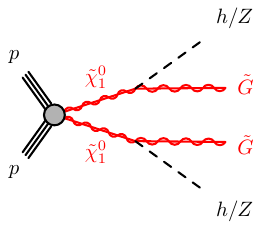}\label{fig:feyn_diag_e}
}
\subfigure[]{
\includegraphics[width=0.28\textwidth]{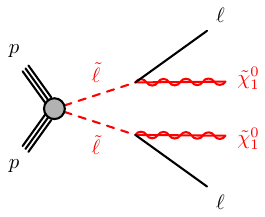}\label{fig:feyn_diag_f}
}

\end{center}
\caption{Simplified models of (a) $\chinoonepm\ninotwo$ production followed by $\chinoonepm\rightarrow \Wpm\ninoone, \ninotwo\rightarrow h/Z \ninoone$; (b) $\ninoone\ninoone$ production followed by $\ninoone\rightarrow h/Z\tilde{G}$; (c) direct slepton pair production followed by $\slepton \rightarrow \ell \ninoone$. No distinction between particle and antiparticle is made in these diagrams.}
\label{fig:feynDiag_weak}
\end{figure}

The electroweak interaction couples SUSY particles to the SM quarks in the protons with a strength that depends specifically on the model parameters affecting the electroweak sector. Therefore, the production cross-section for electroweakinos (Figures~\ref{fig:feyn_diag_d} and \ref{fig:feyn_diag_e}), for example, depends on the mixing in terms of wino, higgsino and (for neutralinos) bino states. Thus the production cross-section for a $\chinoonepm\ninotwo$ pair is larger if each is a pure wino state, and smaller if each is a pure higgsino state, as shown in Figure~\ref{fig:xsec-plot}. Likewise, the production cross-section for a $\ninoone\ninoone$ pair vanishes if the $\ninoone$ is a pure higgsino. Similarly, the production cross-section for slepton pairs (Figure~\ref{fig:feyn_diag_f}) depends on the weak-isospin structure of the sleptons, i.e.\ on the mixing of the slepton in terms of right- and left-handed chirality.

Because of the number of mass eigenstates available and the variety of possible electroweak parameter-space configurations, families of simplified models of electroweakino production have been defined. They stem from specific electroweakino mass hierarchies that arise in limit cases of the electroweak-sector mass parameters. In the MSSM, the soft SUSY-breaking mass term for the bino, winos and higgsinos are indicated with $M_1$, $M_2$ and $\mu$, respectively. The structure of the electroweakino mass matrices is such that a well-defined hierarchy of these parameters translates into a clear mass-eigenstate structure. If, for example, $M_1 < M_2 \ll \mu$, one has a bino-like $\ninoone$ LSP with $\mass{\ninoone} \sim M_1$, and nearly degenerate $\ninotwo$ and $\chinoonepm$ with masses of the order of $M_2$. If $\mu \ll M_1, M_2$, then the first four electroweakinos ($\ninoone$, $\ninotwo$ and two charginos $\chinoonepm$, $\chinotwopm$) will all have almost degenerate masses of the order of $\mu$, leading to a very compressed scenario. Such a higgsino LSP case will yield event topologies with soft decay products emitted in transitions between the electroweakino states. Similarly, $M_2 \ll \mu, M_1$ (wino LSP) yields $\mass{\ninoone} \sim \mass{\chinoonepm} \sim M_2$. Most of the SUSY scenarios that inspire the models described  in Sections~\ref{sec:Weak} and \ref{sec:LLP} refer to one of these paradigms.


%
\section{The ATLAS detector}
\label{sec:atlasdet}
%

%

The ATLAS experiment~\cite{PERF-2007-01} is a multipurpose particle detector with a forward--backward
symmetric cylindrical geometry and nearly \(4\pi\) coverage in solid angle.\footnote{%
ATLAS uses a right-handed coordinate system with its origin at the nominal interaction point (IP)
in the center of the detector and the \(z\)-axis along the beam pipe.
The \(x\)-axis points from the IP to the center of the LHC ring,
and the \(y\)-axis points upwards.
Cylindrical coordinates \((r,\phi)\) are used in the transverse plane,
\(\phi\) being the azimuthal angle around the \(z\)-axis.
The pseudorapidity is defined in terms of the polar angle \(\theta\) as \(\eta = -\ln \tan(\theta/2)\).
Angular distance is measured in units of \(\Delta R \equiv \sqrt{(\Delta\eta)^{2} + (\Delta\phi)^{2}}\).}
It is designed to identify a wide variety of particles and measure their momenta and energies.
These particles include electrons, muons, $\tau$-leptons and photons, as well as gluons and quarks,
which produce collimated jets of particles in the detector.
It consists of an inner tracking detector surrounded by a thin superconducting solenoid
providing a \qty{2}{\tesla} axial magnetic field to measure charged-particle trajectories and momenta, followed by
electromagnetic (EM) and hadron calorimeters that are used in the identification of particles and in the measurement of their energies,
and a muon spectrometer (MS) for measuring the trajectories and momenta of muons.
A two-level trigger system is used to select events~\cite{TRIG-2016-01}. The first-level trigger is implemented in hardware and uses a
subset of the detector information to accept events at a rate below \qty{100}{\kHz} from the LHC's \qty{40}{\MHz} proton bunch crossings.
This is followed by a software-based trigger that reduces the accepted event rate to \qty{1}{\kHz} on average
depending on the data-taking conditions.
An extensive software suite~\cite{ATL-SOFT-PUB-2021-001} is used in data simulation, in the reconstruction
and analysis of real and simulated data, in detector operations, and in the trigger and data acquisition
systems of the experiment.


\section{Analysis strategy}
\label{sec:strategy}

%

The analyses that produced the SUSY mass exclusion limits discussed in Sections~\ref{sec:Strong} to~\ref{sec:LLP} were designed to be complementary, and target  many different event topologies that can arise from the production and decay of SUSY particles. They employ a variety of tools and techniques, which makes each of them largely unique in terms of analysis strategy and, of course, sensitivity to the SUSY parameter space. However, all the analyses discussed apply similar strategies for background estimation, and they all share a common statistical approach in testing, first of all, for compatibility with the SM background prediction (and, therefore, whether the analysis observed a SUSY signal or not), and, in absence of a signal, in setting limits on SUSY production cross-sections, and eventually on sparticle masses.

The design of the analysis starts from the definition of one or more selections determining the signal region(s) (SR). Typically, the SR are designed by maximising the significance of the targeted signal process relative to the SM background as predicted by a Monte Carlo (MC) simulation. Actual collision data are not used in regions with a large expected signal contribution, to avoid any bias while defining the selection. This first phase identifies the set of variables to be used for the selection. They are typically employed in a multivariate analysis, using either a boosted decision tree (BDT) or a neural network (NN), or varied to identify a set of selection criteria to be applied to each of them, sometimes referred to as a \enquote{cut-and-count} approach. The full background estimation procedure often includes the use of data in control regions as described below: the final background level in the SR might be different from that predicted by the MC: the final selection criteria (for the BDT or NN discriminant, or for the individual variables) are therefore adjusted after the final background estimation is in place.

The preliminary SR definition allows the identification of the relevant SM background processes for the analysis. {\em Reducible} background processes pass the signal region selection despite not having the signal characteristics at particle level. Reducible backgrounds are relevant when some events from a high-cross-section SM process (e.g.\ multijet, \Wjets, or \Zjets production) \enquote{fake} one or several of the SR criteria. The faking probabilities are typically small, but the high cross-section may make the process relevant. Reducible backgrounds are often estimated by using techniques that strongly rely on the use of data from real collisions, since the faking mechanisms are not always reproduced reliably by the detector simulation. Some of the most common contexts where estimation of reducible background is relevant are as follows:

\begin{itemize}
\item In analyses where signal events have many leptons, or lepton pairs of the same charge (\enquote{same-sign}, or SS leptons), backgrounds contributions can become relevant,  from jets faking leptons, \enquote{fake} leptons from photon conversions, and non-prompt leptons from hadron decays. Contributions from fake and non-prompt (FNP) leptons are typically estimated using a loose-to-tight method~\cite{EGAM-2019-01}: the FNP contribution to the \enquote{tight} leptons used in the selection is estimated from \enquote{loose} leptons by using scale factors estimated in a dedicated FNP-lepton-dominated selection. Similar techniques are employed to estimate fake-$\tau$-lepton contributions to SR selecting events containing hadronic $\tau$-lepton decays~\cite{ATL-PHYS-PUB-2019-039}.
\item Analyses using SS leptons may be subject to backgrounds arising from processes involving leptons with opposite charges (\enquote{opposite-sign}, or OS leptons), where one of the two lepton charges is misidentified. The charge-flip probability is estimated by comparing yields in OS and SS regions where a $Z\rightarrow \ell \ell$ peak is clearly visible. Obviously, $Z\rightarrow \ell^{\pm}\ell^{\pm }$ arises from $Z\rightarrow \ell^{\pm} \ell^{\mp}$ with a charge-flip. Charge-flip probabilities are typically negligible for muons.
\item The background affecting analyses looking for signals from long-lived particles is almost always reducible. For example, one of the main backgrounds when looking for high-track-multiplicity displaced secondary vertices from the decays of long-lived particles arises from random crossings of tracks that the detector is not able to resolve: the strategy is to determine the probability to have a random crossing from a set of dedicated control regions, and then apply it to events with lower track-multiplicity vertices to estimate the background contribution.
\end{itemize}

\begin{figure}[t]
\begin{center}
\includegraphics[width=1.\textwidth]{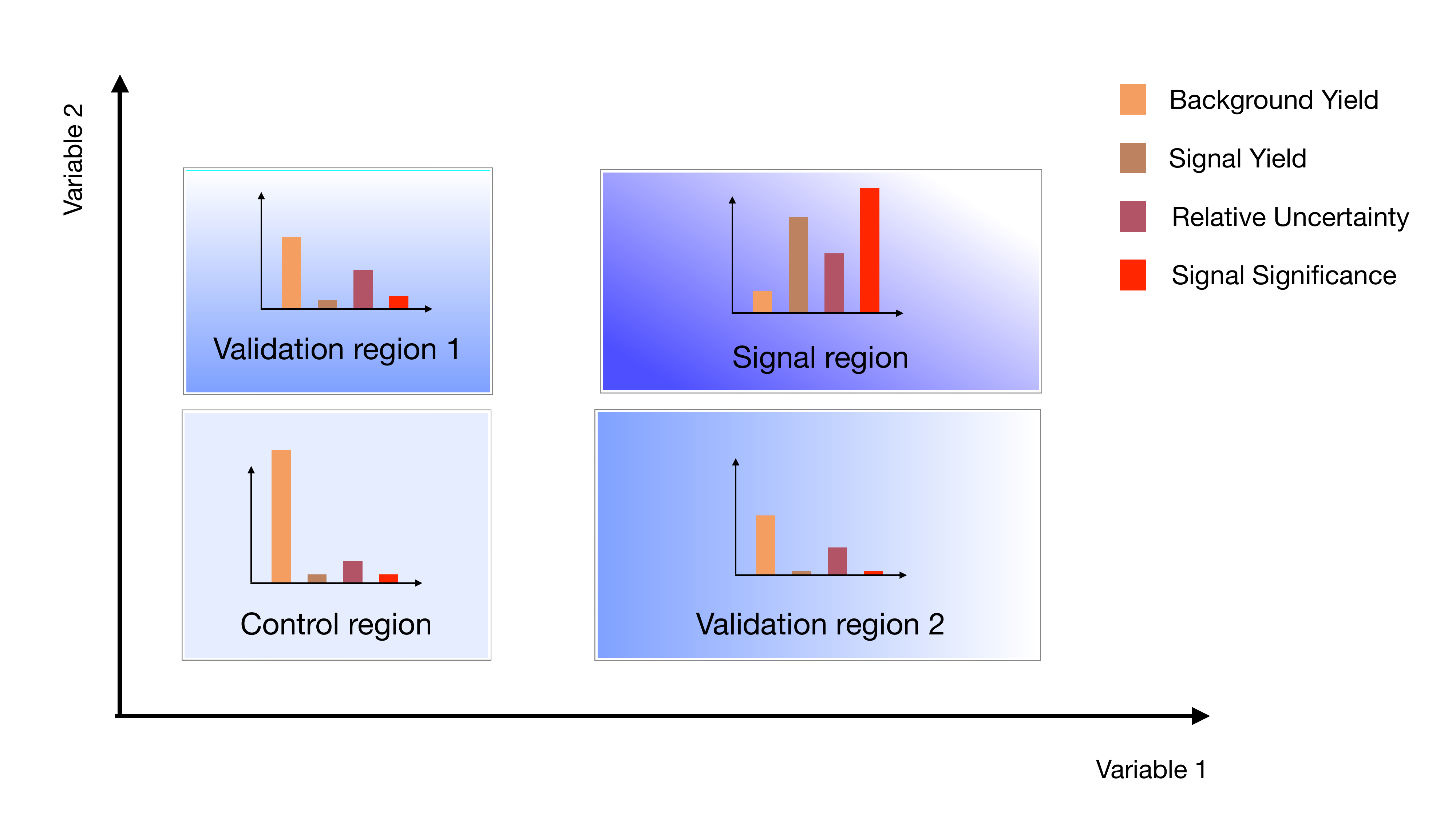}
\end{center}
\caption{Illustrative set-up for an analysis using two variables as the final discriminant. The definition of mutually exclusive signal, control and validation regions is sketched, and indicative values for the signal and background yields, relative uncertainty and signal significance in each region are shown (on an arbitrary scale).}
\label{fig:analysis_strategy}
\end{figure}

{\em Irreducible} backgrounds are those that have the same event topology as the signal. They are typically estimated by relying on MC predictions for their kinematic distributions. Sometimes the overall normalisation of the process is determined in a fit with the help of dedicated control regions (CR), i.e.\ regions where event rates are dominated by the process whose normalisation needs to be estimated. In addition, validation regions (VR) can be defined, with a topology as similar as possible to the SR, but with only a small expected signal contamination: they are used to verify that the background prediction agrees with the data before looking at the yields in the SR. Validation regions are not used in the statistical interpretation of the results. A sketch of a possible generic analysis set-up for an analysis using two variables (for example, two output nodes of a multiclass NN classifier) as the final discriminant, highlighting generic expectations for the signal and background yields, relative background (statistical and systematic) uncertainty, and signal significance, is shown in Figure~\ref{fig:analysis_strategy}.

For the interpretation of the results, a likelihood function is built as the product of Poisson probability functions $P\!\left(n^{\mathrm{obs}}_{i}, n^{\mathrm{exp}}_{i}\right)$, describing the probability of observing $n^{\mathrm{obs}}_{i}$ events in region $i$ (where $i$ is any CR or SR) when $n^{\mathrm{exp}}_{i}$ are expected.  Here $n^{\mathrm{exp}}_{i}$ depends on free-floating normalisation parameters for the signal $\mu_{\mathrm{s}}$ and for the relevant irreducible background processes $\boldsymbol{\mu}_{\mathrm{b}}$. These free-floating parameters are defined so that the yield corresponding to the nominal cross-section is obtained for $\mu = 1$. Systematic uncertainties are treated as nuisance parameters and are constrained with normal distributions $G\!\left(\theta^0_j-\theta_j\right)$, where $\theta^0_j$ is the expected value of the nuisance parameter corresponding to the $j^{\mathrm{th}}$ systematic uncertainty. The parameters $\mu_{\mathrm{s}}, \boldsymbol{\mu}_{\mathrm{b}}$ and $\boldsymbol{\theta}$ are determined by maximum-likelihood fits. The resulting likelihood is:
\begin{align*}
\mathcal{L}\left(\mathbf{n}^{\mathrm{obs}}, \boldsymbol{\theta}^0 | \mu_{\mathrm{s}}, \boldsymbol{\mu}_{\mathrm{b}}, \boldsymbol{\theta}\right) = \prod_{i\,\in \mathrm{\ SR,CR}} P\!\left(n^{\mathrm{obs}}_{i}, n^{\mathrm{exp}}_{i}\!\left(\mu_{\mathrm{s}}, \boldsymbol{\mu}_{\mathrm{b}}, \boldsymbol{\theta}^0, \boldsymbol{\theta}\right)\right) \times \prod_{j\,\in\mathrm{\ syst.\ unc.}} G\!\left(\theta^0_j-\theta_j\right).
\end{align*}
The test statistic used to test whether new physics is observed or to determine exclusion limits is the profile likelihood ratio~\cite{Cowan:2010js}. No significant excess above the SM expectation has been observed at the LHC so far, so this paper concentrates on the resulting exclusion limits. A quantity $\hat{\mu}_{\mathrm{s}}$ is defined as the $\mu_{\mathrm{s}}$ value corresponding to a \cls~\cite{Read:2002hq} of 0.05 for the profile-likelihood-ratio test statistic obtained under the signal-plus-background hypothesis. Models for which $\hat{\mu}_{\mathrm{s}} \le 1$ are said to be excluded at  95\% confidence level (CL). All limits described in this paper are at 95\% CL.


%
\section{Strongly produced supersymmetric particles}
\label{sec:Strong}
%

%

Squarks and gluinos are produced with couplings that are proportional to the strong coupling constant \alphas. Therefore, their production cross-sections are significantly larger than for sleptons and electroweakinos of the same mass. This also implies that squark and gluino production were the first sparticle production processes to which ATLAS achieved sensitivity: for example, gluino were probed up to masses well in the $\TeV$ range already by the end of LHC \RunOne.

\subsection{Gluino pair production}

\begin{figure}[t]
\centering
\includegraphics[width=0.6\textwidth]{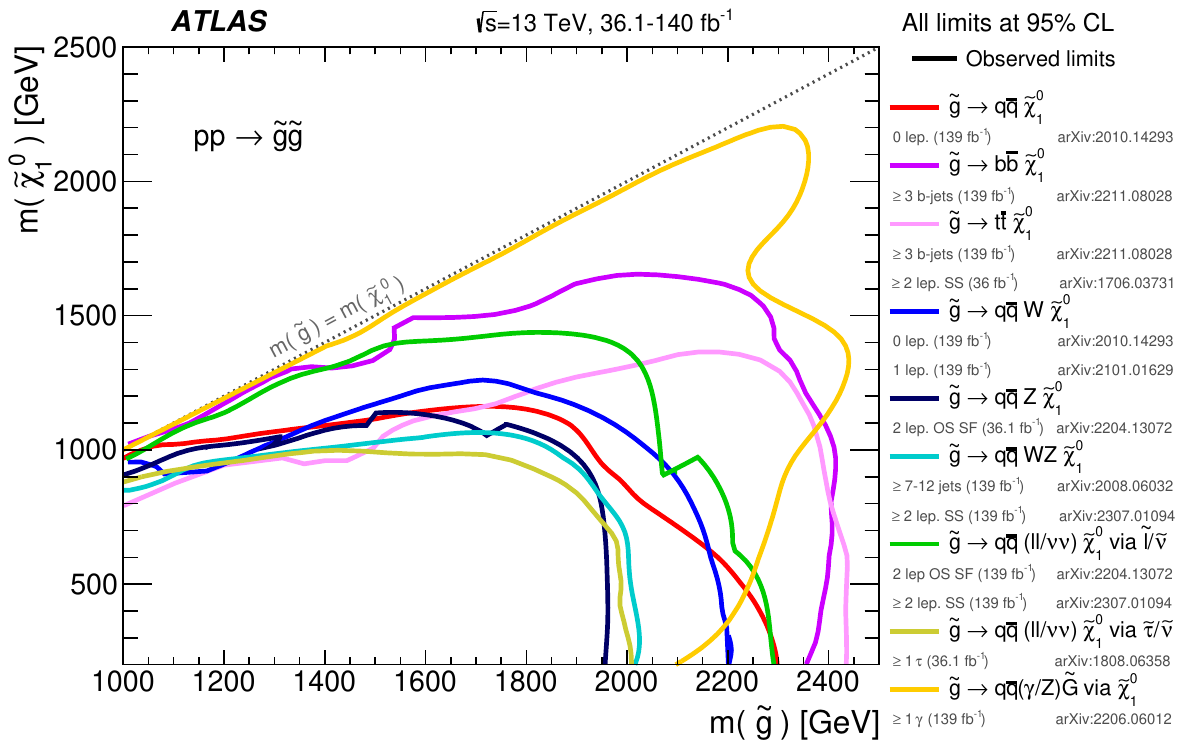}
\caption{Exclusion limits from $13\,\TeV$ data in the $\mass{\gluino}$--$\mass{\ninoone}$ plane for different simplified models featuring the decay of the gluino to the lightest supersymmetric particle (lightest neutralino or gravitino) either directly or through a cascade chain featuring other SUSY particles with intermediate masses. For each line, the gluino decay mode is reported in the legend and it is assumed to proceed with 100\% branching fraction. Some limits depend on additional assumptions about the masses of intermediate states, as described in the references cited in the plot. Where more than one reference is given for each line, the total exclusion limit from multiple analyses is shown.
}
\label{fig:GG_Direct}
\end{figure}

The sensitivity of  ATLAS to gluino pair production followed by RPC decays of the $\gluino$ is shown in Figure~\ref{fig:GG_Direct}. All the limits shown were obtained by assuming a simplified signal model and the cross-sections shown in Figure~\ref{fig:xsec-plot}: in particular, the production cross-section is a function of the gluino mass only, and it therefore decreases strongly at increasing values along the $\mass{\gluino}$-axis. In all cases, the limits are presented as a function of the gluino mass and lightest neutralino mass. Different curves represent either different assumptions about the gluino decay chain, or different analysis results.

Focusing on the limit intercepts with the $\mass{\gluino}$-axis, i.e.\ those for $\mass{\ninoone} = 0\ \GeV$, the first observation is that the $2\,\TeV$ range predicted from simple cross-section considerations in Section~\ref{sec:models} is indeed realised. The second observation is that the limits on the gluino mass obtained with dedicated optimised analyses depend significantly on the assumed gluino decay, and on $\mass{\ninoone}$. This is readily understood as a consequence of the total energy available in the decay being determined by the mass difference $\mass{\gluino} - \mass{\ninoone}$. A small value of this quantity corresponds to smaller values of the momenta of the final-state objects, and, therefore, to signal topologies that are more difficult to separate from the SM background.

The different models in Figure~\ref{fig:GG_Direct} correspond to scenarios that assume different RPC gluino decays. Together, they give a very good overview of the extent of the effort made in the search for this sparticle.

The simplest RPC Feynman vertex that can be considered is one where the gluino couples to a squark and a quark. The flavour of the squark, and the way it couples to the electroweak sector determines the complexity of the final state. In one of the simplest simplified models, the squarks and all electroweakino states but one ($\ninoone$) have a very large mass. The first case considered is one where the squark is not the stop. The gluino decay chain is $\gluino \rightarrow \squark^* q \rightarrow qq\ninoone$. The case where the virtuality of the $\squark^*$ is such that the $\gluino$ becomes long-lived is phenomenologically compelling and theoretically very interesting (arising from models of split-SUSY~\cite{Giudice:2004tc,ArkaniHamed:2004fb}), and it is discussed in Section~\ref{sec:LLP}. Here the discussion is limited to the case where the $\gluino$ decay is prompt. The analysis that drives the sensitivity~\cite{SUSY-2018-22} to this scenario targets both direct gluino and direct squark pair production by requiring no leptons in the final state and significant \MET, which is also exploited for triggering purposes: the corresponding limit is indicated by the red line in Figure~\ref{fig:GG_Direct}. The analysis selection is based on the jet multiplicity in the event (\Njet), the significance of the missing transverse momentum \metsigHT, and $\meff = \HT + \MET$. Here \HT denotes the scalar sum of the transverse momenta (\pt) of the jets in the event. The dominant background process in the considered signal regions is $Z(\rightarrow \nu\bar{\nu})\,+$\,jets. This is estimated using simulated events corrected with auxiliary measurements of the $\gamjets$ and $Z(\rightarrow\ell^{\pm}\ell^{\mp})\,+$\,jets processes in a similar phase-space region. Other relevant background processes are $\Wjets$ and $\ttbar$ production, with a $W\rightarrow\ell \nu$ decay to satisfy the \MET requirement. For the latter class of processes, the lepton either fails the identification requirements or is a hadronically decaying $\tau$-lepton.
A small contribution from multijet production is estimated with the jet smearing method, where the background prediction is seeded from simulated events corrected to account for the detector's response to jets as measured in data~\cite{SUSY-2011-20}. Gluino masses up to about $2.2\,\TeV$ are excluded for massless $\ninoone$. For gluinos decaying to $\ninoone$ LSPs, there is no sensitivity to gluinos with masses above about $1.3\,\TeV$ if the masses of the gluino and $\ninoone$ are similar.  It is important to always bear in mind that the obtained limits assume a branching fraction of 100\% for the particular decay. Also, the assumption that squarks do not affect the gluino production cross-sections can be quite a strong one, requiring squarks with masses of tens of $\TeV$. Further interpretations in Ref.~\cite{SUSY-2018-22} address some of these points, providing limits in models where the gluino and squark masses are similar.

\begin{figure}[t]
\centering
\includegraphics[width=0.6\textwidth]{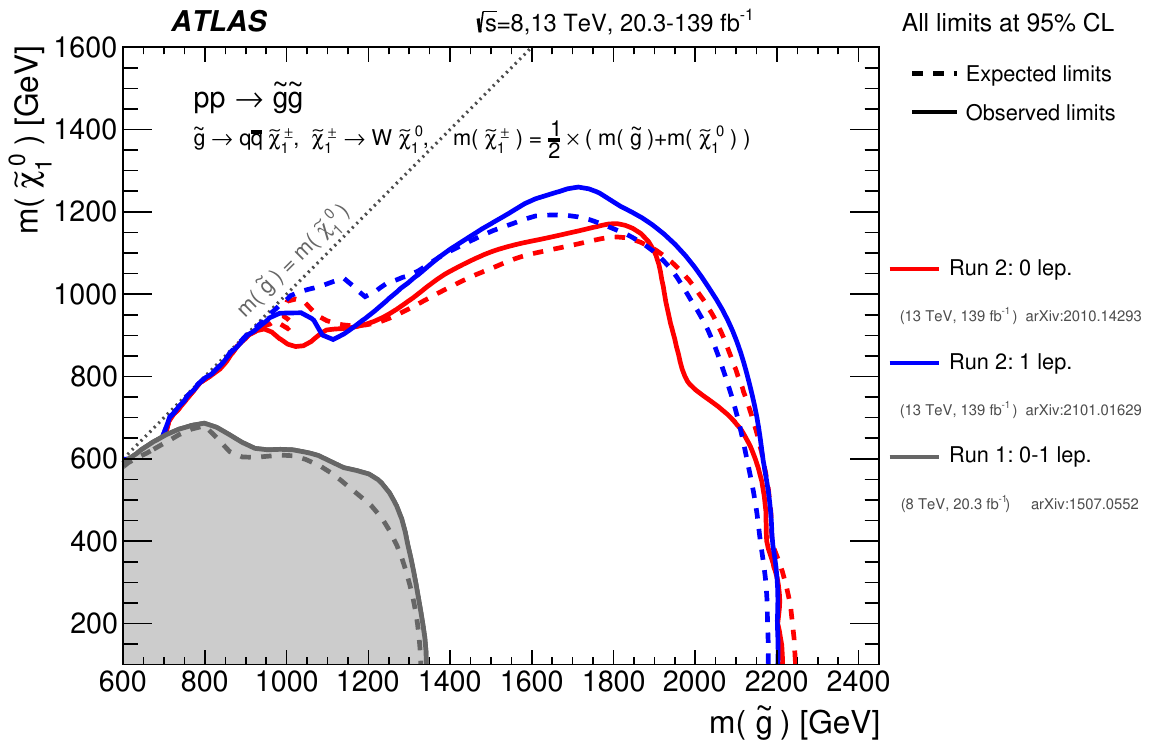}
\caption{Exclusion limits from $13\,\TeV$ data in the $\mass{\gluino}$--$\mass{\ninoone}$ plane for the simplified model where a pair of gluinos are produced, and each decays promptly via an on-shell chargino into a pair of quarks, a $W$ boson, and the lightest neutralino. The chargino mass is assumed to be midway between the gluino and neutralino masses. Theoretical signal cross-section uncertainties are not included in the limits shown.}
\label{fig:GG_1step}
\end{figure}

If the assumptions about the electroweak sector are modified, and a richer spectrum of relatively light electroweakino states is allowed, then the phenomenology becomes more complex, with different squark decays potentially competing with each other. ATLAS has considered various benchmark gluino decays, especially focusing on models involving longer gluino decay chains. In some cases, the gluino decay chain may lead to the production of vector bosons, which opens up possibilities for final states containing leptons, possibly with same-sign charges. Additionally, longer gluino decay chains with hadronic vector-boson decays can result in final states with a higher jet multiplicity or involve hadronic resonances. Figure~\ref{fig:GG_Direct} shows some of the limits obtained in these scenarios in different shades of blue. In one scenario (light blue line in Figure~\ref{fig:GG_Direct}, also shown in Figure~\ref{fig:GG_1step}), the squark is assumed to decay as $\squark^* \rightarrow q\chinoonepm \rightarrow qW\ninoone$, leading to $\gluino \rightarrow q\bar{q}' W \ninoone$. This scenario is targeted both by Ref.~\cite{SUSY-2018-22}, already discussed, and by an analysis requiring the presence of one lepton from the $W$ decay~\cite{SUSY-2018-10}. The latter applies a selection that rejects background processes containing one $W$ boson decaying leptonically by requiring large values of the transverse mass \mt of the lepton and $\ptmiss$. The selection is optimised for different  the jet multiplicities. The rest of the selection exploits the topology of the expected final state by using variables such as  \MET, \Njet, and \meff (including the lepton \pt in the scalar sum of transverse momenta). The two analyses selecting events with either no or one lepton in the final state have dedicated selections for models where the \gluino and \ninoone masses are similar, resulting in improved exclusion limits for such compressed mass-difference scenarios. The resulting exclusion limits from the two analyses in this simplified model are similar, with each excluding a range up to $\mass{\gluino} = 2.2\,\TeV$ for $\mass{\ninoone} = 0$. The limit is reduced by about $1\,\TeV$ in the compressed scenario.

Other cases considered explicitly in Figure~\ref{fig:GG_Direct} are those with $\squark^*\rightarrow q\ninotwo \rightarrow qZ\ninoone$, targeted by an analysis requiring an opposite-sign same-flavour (OS~SF) lepton pair compatible with a $Z$ boson in association with $\MET$ and jets~\cite{SUSY-2018-05}, and those arising from a longer decay chain such as $\squark^*\rightarrow q\chinoonepm\rightarrow q \ninotwo W \rightarrow qWZ\ninoone$, targeted by multiple analyses requiring either large jet multiplicities~\cite{SUSY-2018-17} or SS leptons in addition to jets~\cite{ATLAS:2023afl}. Scenarios involving intermediate electroweakino decays via sleptons have also been considered.

The yellow line in Figure~\ref{fig:GG_Direct} displays a behaviour completely different from the others. It applies to scenarios inspired by General Gauge Mediation (GGM)~\cite{Meade:2008wd}, where the LSP is often a light gravitino with a mass around $1\,\GeV$ or less. The phenomenology in these scenarios is determined by the nature of the next-to-lightest supersymmetric particle (NLSP). The case shown here is one where the NLSP is a neutralino that decays to the LSP with photon or $Z$ boson emissions, and the resulting topology includes photons in the final state, yielding a distinctive $\gamma + \MET + \jets$ final-state signature. The amount of $\MET$ in the event largely determines the analysis acceptance: in this case it is proportional to $\mass{\ninoone}-\mass{\tilde{G}}$, so the maximum acceptance is obtained for large neutralino masses. Overall, the limits on the gluino mass are in line with those of the other analyses.

The cases where the virtual squark produced in the gluino decay is a third-generation squark are singled out in pink and purple in Figure~\ref{fig:GG_Direct}. The physics reason to single them out is connected with naturalness considerations and is discussed in Section~\ref{sec:squark_production}. The topological reason to single them out comes from the presence of $b$-quarks in the final state, yielding a powerful experimental handle with the application of $b$-tagging techniques~\cite{FTAG-2019-07} to identify \bjets. Neglecting potential flavour-violating decays of the stop or sbottom, the decay of a gluino pair via a third-generation squark with a branching fraction of 100\% yields four \bjets in the final state. An analysis selecting events with three or more $b$-jets~\cite{SUSY-2018-30} dominates the sensitivity to these scenarios. The selection is based on a neural network that uses the kinematic variables of the jets, \bjets, leptons and $\ptmiss$ as input. The network was trained to separate the $\gluino\rightarrow t\bar{t} \ninoone$ and  $\gluino\rightarrow b\bar{b} \ninoone$ decays from the SM background, dominated by \ttbar, single top and \Zjets production. A different selection, which did not make use of multivariate techniques, was optimised to target mixed final states such as those arising from scenarios where both the stop and the sbottom take part in gluino decay, leading to final states with variable numbers of $b$- and $t$-quarks.

\begin{figure}[t]
\centering
\includegraphics[width=0.6\textwidth]{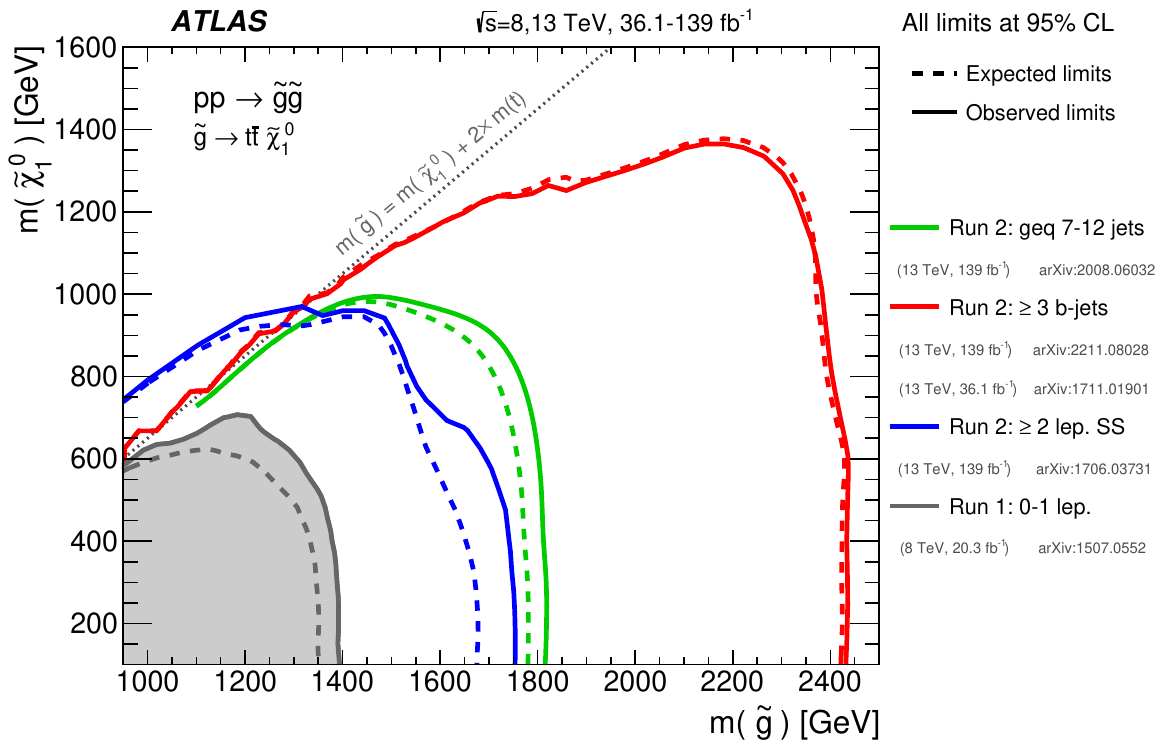}
\caption{Exclusion limits from $13\,\TeV$ data in the $\mass{\gluino}$--$\mass{\ninoone}$ plane for a simplified model where a pair of gluinos decay promptly via off-shell top squarks into four top quarks and two lightest neutralinos. The SS lepton analysis sensitivity covers $m(\gluino) < m(\ninoone) + 2m(t)$ due to the off-shell top quark contribution.  Theoretical signal cross-section uncertainties are not included in the limits shown. Where more than one reference is given for each line, the total exclusion limit from multiple analyses is shown.}
\label{fig:GGtt}
\end{figure}

Summary limits on the model of gluino pair production yielding four top quarks and \MET are shown in Figure~\ref{fig:GGtt}. The richness of the topology of the final state makes it a target of many analyses. In particular, SS analyses achieve sensitivity to regions of the parameter space where the top quarks emitted in the decay are virtual.

\subsection{Squark pair production}
\label{sec:squark_production}

\begin{figure}[t]
\centering
\includegraphics[width=0.6\textwidth]{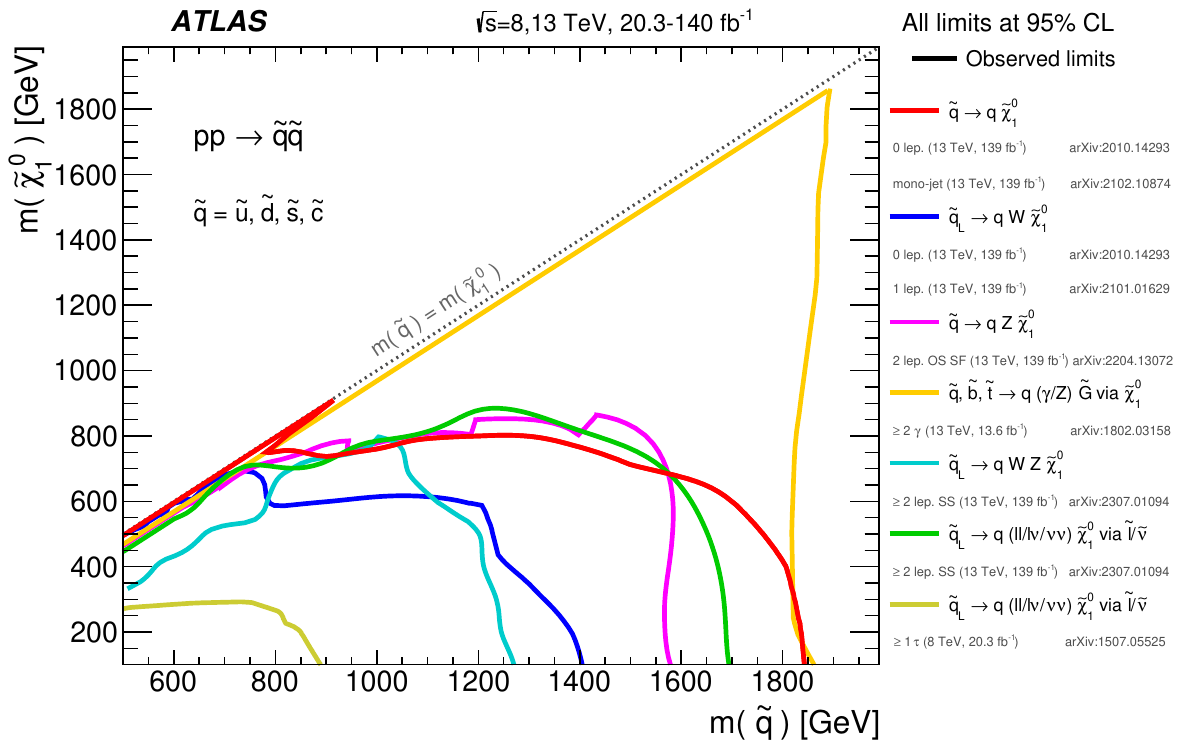}
\caption{Exclusion limits from $8\,\TeV$ and $13\,\TeV$ data in the $\mass{\squark}$--$\mass{\ninoone}$ plane for different simplified models featuring the decay of squarks to the lightest supersymmetric particle (lightest neutralino or gravitino) either directly or through a cascade chain featuring other SUSY particles with intermediate masses. For each line, the squark decay mode is reported in the legend and it is assumed to proceed with 100\% branching fraction. The limits on $\squarkL \rightarrow q W \ninoone$ assume the chargino mass to be midway between the squark and neutralino masses. The other $\squarkL$ interpretations assume the first intermediate particle to be midway between the squark and neutralino masses, and the second intermediate particle to be midway between the first intermediate particle and neutralino masses. The additional assumptions about the masses of the intermediate states are described in the references cited in the plot. Where more than one reference is given for each line, the total exclusion limit from multiple analyses is shown.
}
\label{fig:QQ_Direct}
\end{figure}

The search for pair production of squarks is closely connected with that of gluinos. A summary of the mass exclusion limits is shown in Figure~\ref{fig:QQ_Direct}, assuming an eightfold mass degeneracy for the squarks (two chirality states for each of the four lightest SM flavours). Depending on the structure and mass hierarchy of the electroweak sector, the squarks will have decay chains that resemble those already discussed for the gluino pair production. If the only electroweak state that couples to the quark is a neutralino LSP, then the squark decay chain is $\squark \rightarrow q \ninoone$. The final state will contain jets and $\met$, and the analysis discussed in  Ref.~\cite{SUSY-2018-22} has good sensitivity for most values of the mass difference $\Delta m(\squark, \ninoone)  = \mass{\squark} - \mass{\ninoone}$. For small $\Delta m(\squark, \ninoone)$, the final state contains very small \MET, unless one focuses on the production of squarks recoiling against substantial QCD initial-state radiation (ISR). This case is explicitly targeted in Ref.~\cite{EXOT-2018-06}: the analysis searches for an excess of events with large \MET recoiling against substantial hadronic activity. After a number of selections aimed at making sure that the events under investigation do not arise from jet energy mismeasurement or beam backgrounds, the dominant background processes are $W(\rightarrow \ell \nu)\,+$\,jets (where the $\ell$ is lost or is a hadronically decaying $\tau$-lepton) and $Z(\rightarrow \nu \bar{\nu})\,+$\,jets production. A sophisticated background estimation procedure exploits the similarity of these processes with $Z(\rightarrow \ell^{\pm} \ell^{\mp})\,+$\,jets and $\gamjets$.  As in the gluino case, a richer electroweak spectrum leads to longer decay chains, producing one-step $\squark\rightarrow q W \ninoone$ \cite{SUSY-2018-22, SUSY-2018-10} or $\squark \rightarrow q Z^{(*)} \ninoone$ \cite{SUSY-2018-05} decays, or two-step $\squark \rightarrow qWZ\ninoone$ \cite{ATLAS:2023afl} decays. Similarly to the gluino case, scenarios involving sleptons in the electroweak-state decays, or GGM-like scenarios with a gravitino LSP, were also considered and targeted by specific analyses.

Overall, mass exclusion limits for eightfold-degenerate squarks range from $1.3\,\TeV$ to $1.8{-}1.9\,\TeV$ for $\mass{\ninoone} = 0\,\GeV$, and worsen with decreasing values of $\Delta m(\squark, \ninoone)$. For $\Delta m(\squark, \ninoone)$ of the order of a few $\GeV$, the exclusion limit is about 900\,\GeV. These limits were obtained assuming simplified models.

The production cross-section for squarks scales linearly with the assumed degeneracy.  For example, the limit in Ref.~\cite{SUSY-2018-22} for $\mass{\ninoone} = 0\,\GeV$ decreases from about $1.85\,\TeV$ to about $1.2\,\TeV$ if the production of a single \squark chirality state is assumed.

Third-generation squarks ($\sbottom$ and $\stop$), and to some extent the $\tilde{c}$, have been targeted by dedicated analyses. The presence of $b$-jets (or $c$-jets) in the final state offers an additional experimental handle for improving the background rejection. The third-generation squarks are also constrained to be at the $\TeV$ scale by naturalness considerations: because of the large Yukawa coupling, stop loops are the main contribution to the first-order corrections to the lightest Higgs boson's mass in the MSSM~\cite{Heinemeyer:2004ms}. Naturalness requirements place constraints on the stop masses and mixing between $\stopL$ and $\stopR$. They also constrain the $\sbottomL$ because it belongs to the same weak-isospin doublet as the $\stopL$ and therefore shares the same SUSY mass parameter.

\begin{figure}[t]
\centering
\includegraphics[width=0.6\textwidth]{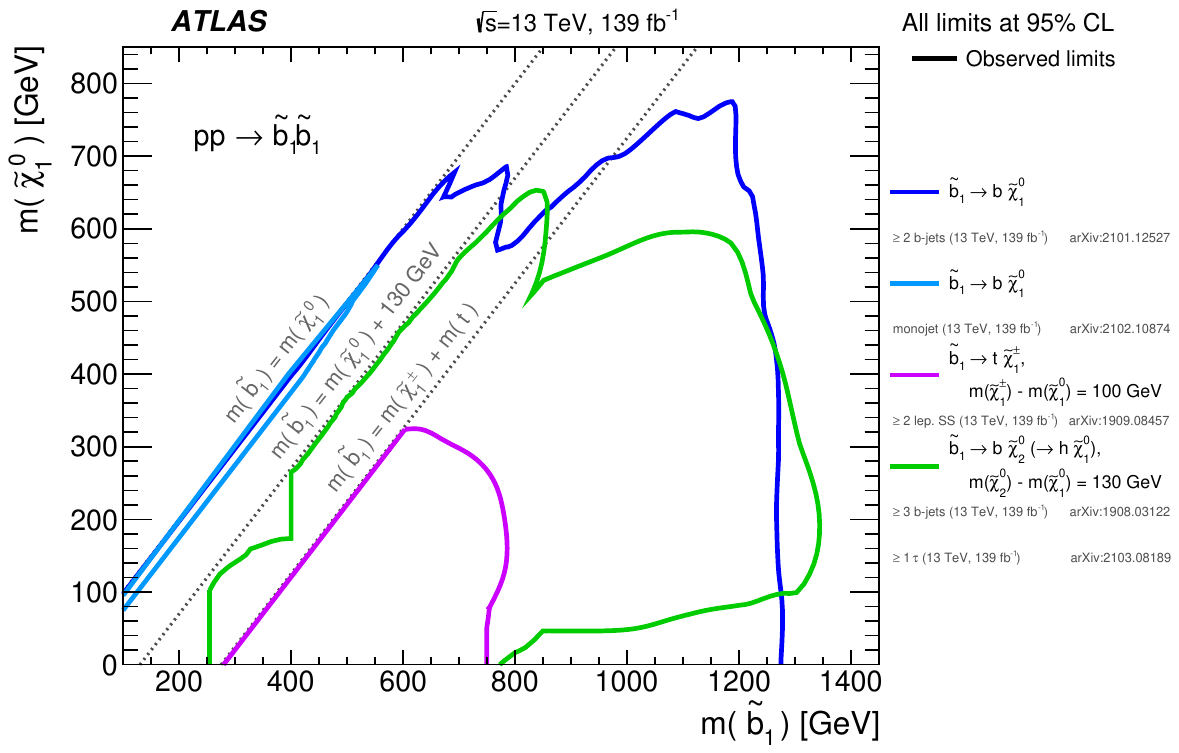}
\caption{Exclusion limits in the $\mass{\sbottomone}$--$\mass{\ninoone}$ plane for direct sbottom production. The lightest neutralino ($\ninoone$) is assumed to be the LSP. Several different decay scenarios are shown, along with different parameterisations of the intermediate particles in the models. Where more than one reference is given for each line, the total exclusion limit from multiple analyses is shown.}
\label{fig:SbottomPair_Direct}
\end{figure}

A summary of the $\sbottom$ mass exclusion limits is shown in Figure~\ref{fig:SbottomPair_Direct}. The limit in blue corresponds to the simplest decay chain, involving only a $\ninoone$ LSP. The analysis targeting this final state~\cite{SUSY-2018-34} requires the presence of $\bjets$ and \MET, and exploits the presence of kinematical endpoints in contransverse mass~\cite{Tovey:2008ui} for the dominant $\ttbar$ production background. Compressed scenarios are targeted by an analysis exploiting ISR hadronic activity and using dedicated techniques for tagging low-\pt $\bjets$ \cite{ATLAS-CONF-2019-027}. Longer \sbottom decay chains are targeted by analyses explicitly looking for the presence of a Higgs boson in the $\ninotwo\rightarrow h \ninoone$ decay~\cite{SUSY-2018-31,SUSY-2018-40}.

Dedicated training of neural networks using input variables sensitive to the displaced decays of $b$-hadrons can produce relatively efficient $c$-jet tagging algorithms, which can be used effectively in the search for $\tilde{c} \rightarrow c \ninoone$. The analysis~\cite{SUSY-2016-26} was performed with a fraction of the available \RunTwo integrated luminosity and follows a logic similar to that in Ref.~\cite{SUSY-2018-34}.

\begin{figure}[t]
\centering
\includegraphics[width=0.8\textwidth]{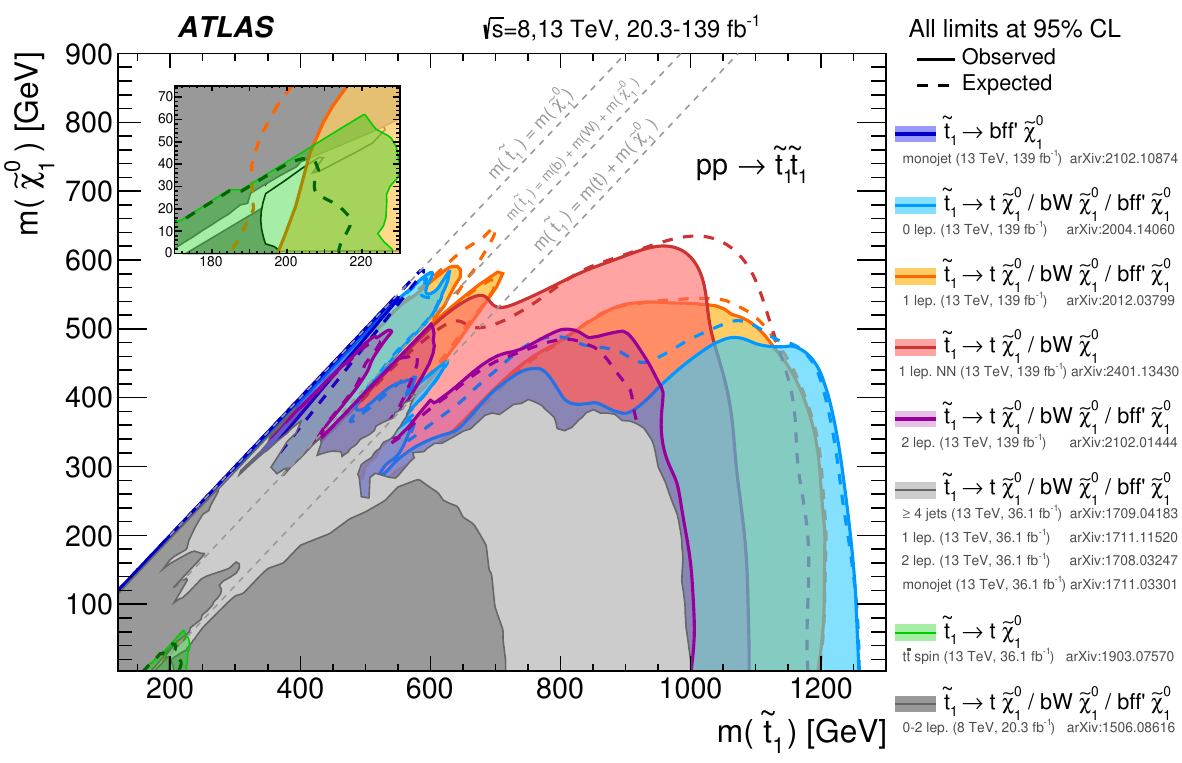}
\caption{Summary of the dedicated ATLAS searches for top squark (stop) pair production based on $pp$ collision data taken at $\sqrt{s} = 13 \,\TeV$. Exclusion limits are shown in the $\mass{\stopone}$--$\mass{\ninoone}$ plane. The dashed and solid lines show the expected and observed limits, respectively, including all uncertainties except the theoretical signal cross-section uncertainty (PDF and scale). Three decay modes are considered separately with 100\% branching fraction: $\stopone \rightarrow t + \ninoone$ (where the $\stopone$ is mostly $\stopR$), $\stopone\rightarrow W b \ninoone$ (three-body decay for $\mass{\stopone} < \mass{t} + \mass{\ninoone}$), and $\stopone \rightarrow  ff'b\ninoone$ (four-body decay). Where more than one reference is given for each line, the total exclusion limit from multiple analyses is shown.}
\label{fig:StopPair_Direct}
\end{figure}

Stop pair production was the target of an extensive search campaign as early as \RunOne~\cite{SUSY-2014-07}, and \RunTwo saw  further development of the search strategy, with a variety of new final states being targeted. The popularity of stop searches is due to the stop's connection with the naturalness problem. From an experimental point of view, the phenomenology of the stop decay is very rich because of the large top quark mass: assuming the simplest electroweak-sector configuration (a $\ninoone$ LSP), the kinematics of the decay is determined not only by the \stopone and \ninoone masses, but also by how their difference $\Delta m(\stopone, \ninoone)$  compares  to the top quark and $W$ boson masses. If  $\Delta m(\stopone, \ninoone) > \mass{t}$, then the $\stopone$ will experience a two-body decay $\stopone\rightarrow t\ninoone$, followed by an on-shell decay of the top quark into an on-shell $W$ boson and a $b$-quark. If $\mass{W} + \mass{b} < \Delta m(\stopone, \ninoone) < \mass{t}$, the stop will decay into three bodies, $\stopone \rightarrow b W \ninoone$, via an off-shell top. Finally, if $\mass{b} < \Delta m(\stopone, \ninoone) < \mass{W} + \mass{b}$, the decay will be into four bodies, $\stopone \rightarrow bf f' \ninoone$, with $f$ representing a generic fermion. These different regions are clearly displayed in Figure~\ref{fig:StopPair_Direct}. The parameter space is targeted with a suite of analyses looking for different lepton (electron and muon) multiplicities. Depending on the decay of the $W^{(*)}$, the final state can contain zero~\cite{SUSY-2018-12}, one~\cite{SUSY-2018-07} or two~\cite{SUSY-2018-08} leptons.

The zero-lepton analysis defines a series of different signal regions based on the number of top quark and $W$ boson decays that can be tagged by making use of large-radius jets to collect all the decay products. Depending on the region, the selection is completed by using additional variables such as \MET and the transverse mass of the \MET and \bjets to reject the dominant background from \ttbar and \Zjets production. As in Ref.~\cite{SUSY-2018-34}, the compressed region is targeted with dedicated tools to tag low-\pt \bjets.

The one-lepton analysis also utilises a set of signal regions targeting the two-, three- and four-body decays separately. A challenging background process  is \ttbar production where both top quarks decay leptonically, and one lepton is lost or is a hadronically decaying $\tau$-lepton. This background process is targeted by making use of a variable called {\em topness} ~\cite{Graesser:2012qy}, which yields an estimate of how much a given event with one lepton in the final state resembles a dileptonic \ttbar decay, based on expected detector resolutions and mass constraints present in such events. An updated result (shown in red in Figure~\ref{fig:StopPair_Direct}) is obtained by separating events into mutually exclusive categories based on the number of reconstructed large-radius jets, and the number of those compatible with arising from the decay of a top quark: for each category, dedicated neural networks are trained to disentangle the signal from the background. This improves the sensitivity, especially at intermediate stop masses and large neutralino masses.

A challenging mass hierarchy is the one where $\Delta m(\stopone,\ninoone) \approx \mass{t}$: the final state and kinematics of $\stopone\stopone\rightarrow  t\bar{t}\ninoone\ninoone$ closely resemble those from \ttbar production, especially if $\mass{\ninoone} \approx 0$. In this case, precision measurements of SM \ttbar production can come to the rescue, as indicated in the inset plot of Figure~\ref{fig:StopPair_Direct}:  a precise measurement of spin correlations in \ttbar events~\cite{TOPQ-2016-10} allows constraints to be placed on the additional stop pair production.

Figure~\ref{fig:StopPair_Direct} shows that stop pair production is excluded (for the decays considered) up to stop masses of about $1.25\,\TeV$ for $\mass{\ninoone} = 0$. The limit drops to about $\mass{\stopone} \sim 600\ \GeV$ in the compressed region.

If a more complex electroweak sector is considered, for example one allowing $\stopone\rightarrow b \chinoonep$ and $\stopone \rightarrow t \ninotwo$, then additional kinematic configurations and particles become possible. Typically, the analyses discussed have good sensitivity to these scenarios. Additional analyses explicitly looking for the presence of a Higgs or $Z$ boson in the final state have also been developed~\cite{SUSY-2018-21}, achieving interesting sensitivities. The case of a stop decay explicitly involving a $\stau$ has also been explored~\cite{SUSY-2019-18}. Finally, the flavour-changing decay of a stop into a charm quark and a neutralino LSP was targeted in Ref.~\cite{ATLAS:2024vyj}.


%
\section{Weakly produced supersymmetric particles}
\label{sec:Weak}
%

%

The electroweak production of SUSY particles leads to cross sections that are orders of magnitude smaller than those for strong production (for a given sparticle mass - see Figure~\ref{fig:xsec-plot}), with final-state signatures that are difficult to separate from SM processes.
The ATLAS search programme aims to exploit all channels -- from the clean, but rare, leptonic final states to the higher branching-fraction hadronic decays -- in the quest to discover SUSY.

\subsection{Slepton pair production}

First- and second-generation slepton production, with decays into a charged lepton and a $\ninoone$ LSP ($\slepton\slepton$, with $\slepton \rightarrow \ell\ninoone$) with 100\% branching fraction, results in a clean final state of two OS~SF leptons ($e^+e^-$ or $\mu^+\mu^-$) and $\met$.
A \RunTwo search~\cite{SUSY-2018-32} considering general slepton production scenarios with large mass splittings between the $\slepton$ and the $\ninoone$ improved the  sensitivity to higher slepton and neutralino masses with respect to  \RunOne searches.
A more recent search was more targeted and focused on the more compressed slepton production scenarios with $m(\slepton)-m(\ninoone)<100\,\GeV$~\cite{SUSY-2019-02}, dividing the search between signatures with or without an ISR jet to exploit the boosted topology of $\slepton\slepton$ production with ISR.
Both approaches use multiple bins in stransverse mass, $\mttwo$~\cite{Lester:1999tx}, which are optimised for mass exclusion potential in $\slepton\slepton$ production.
SM backgrounds containing $Z$ boson or low-mass resonances are suppressed by discarding OS~SF lepton pairs with compatible invariant mass, while $\ttbar$ production is suppressed by discarding events containing $b$-tagged jets.
At least moderate $\met$ significance is required in order to suppress backgrounds with misreconstructed $\met$.

The most compressed slepton scenarios are targeted using a tailored analysis requiring two low-\pT leptons, an ISR jet, and high $\met$~\cite{SUSY-2018-16}.
Low $\mttwo$ is used for signal--background discrimination, along with $R_{\mathrm{ISR}}$, the ratio of the $\met$ to the $\pt$ of the ISR system.
A novel muon identification working point is used to recover some of the efficiency otherwise lost as $3\,\GeV<\pt<6\,\GeV$ muons traverse the calorimeters.

\begin{figure}[t]
\centering
\includegraphics[width=0.6\textwidth]{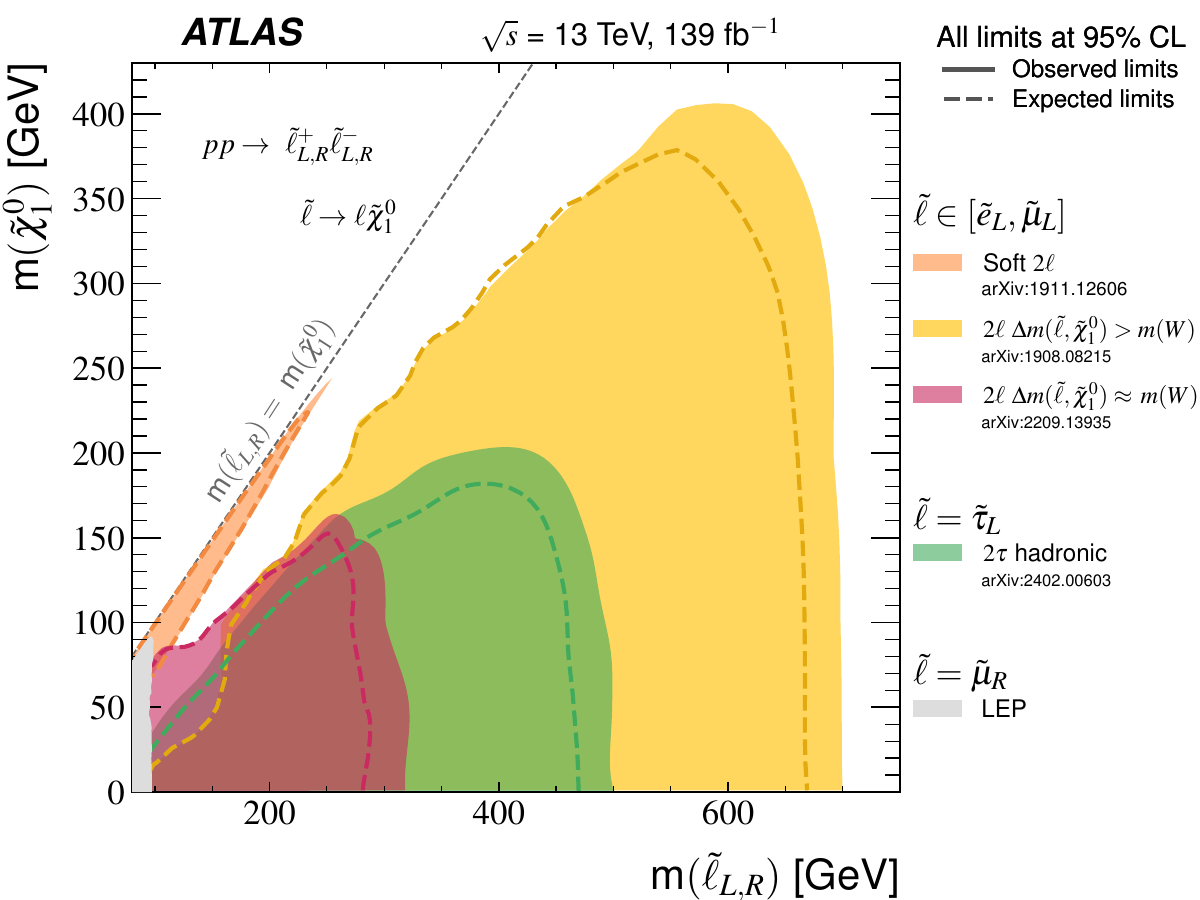}
\caption{Exclusion limits in the $\mass{\slepton}$--$\mass{\ninoone}$ plane for different analyses probing the direct production of sleptons with decays into a lepton and neutralino. The type of sleptons (flavour and coupling) included in each search is specified in the legend.
}
\label{fig:SleptonPair_Direct}
\end{figure}

The three searches see no significant excesses in the data compared to the SM background expectation, allowing limits to be set in simplified SUSY models, as shown in Figure~\ref{fig:SleptonPair_Direct}.
For mass-degenerate left- and right-handed slepton pair production, $\slepton$ masses up to $700\,\GeV$ are excluded for massless $\ninoone$.
For the very compressed scenarios with $\mass{\slepton}-\mass{\ninoone} \sim 10\,\GeV$, sleptons are excluded up to masses of $250\ \GeV$.
The so-called \enquote{slepton-gap} between the LEP and LHC limits is now closed with the results from the latest slepton search, excluding sleptons up to $150\ \GeV$ for $\mass{\slepton}-\mass{\ninoone}\sim 50\,\GeV$. This still leaves an interesting parameter-space region to explore between the sensitivities in Ref.~\cite{SUSY-2018-16} and Refs.~\cite{SUSY-2018-32,SUSY-2019-02}.

Searches for third-generations sleptons, the staus, use final states with two hadronically decaying $\tau$-leptons and high $\met$.
The analysis in Ref.~\cite{SUSY-2018-04} uses a cut-and-count approach with two signal regions.
The more recent analysis in Ref.~\cite{SUSY-2023-03} %
improves upon this by using four overlapping BDTs targeting different $\stau$--$\ninoone$ mass regimes.
All four BDT signal regions show a $0.7\sigma{-}1.3\sigma$ deficit -- a common deficit since the signal regions partially overlap.
The exclusion limits for mass-degenerate left- and right-handed staus are shown in Figure~\ref{fig:SleptonPair_Direct}, where stau masses up to $475\,\GeV$ are excluded.
When considering the production of purely right-handed staus, which has a lower cross-section, staus masses up to $350\,\GeV$ are excluded (not shown in the plot).
There is little sensitivity to the more compressed $\stau$--$\ninoone$ mass scenarios, because the searches rely on high-$\pt$ hadronically decaying $\tau$-leptons to trigger the event.

\subsection{Electroweakino pair production}

\begin{figure}[t]
\centering
\includegraphics[width=0.6\textwidth]{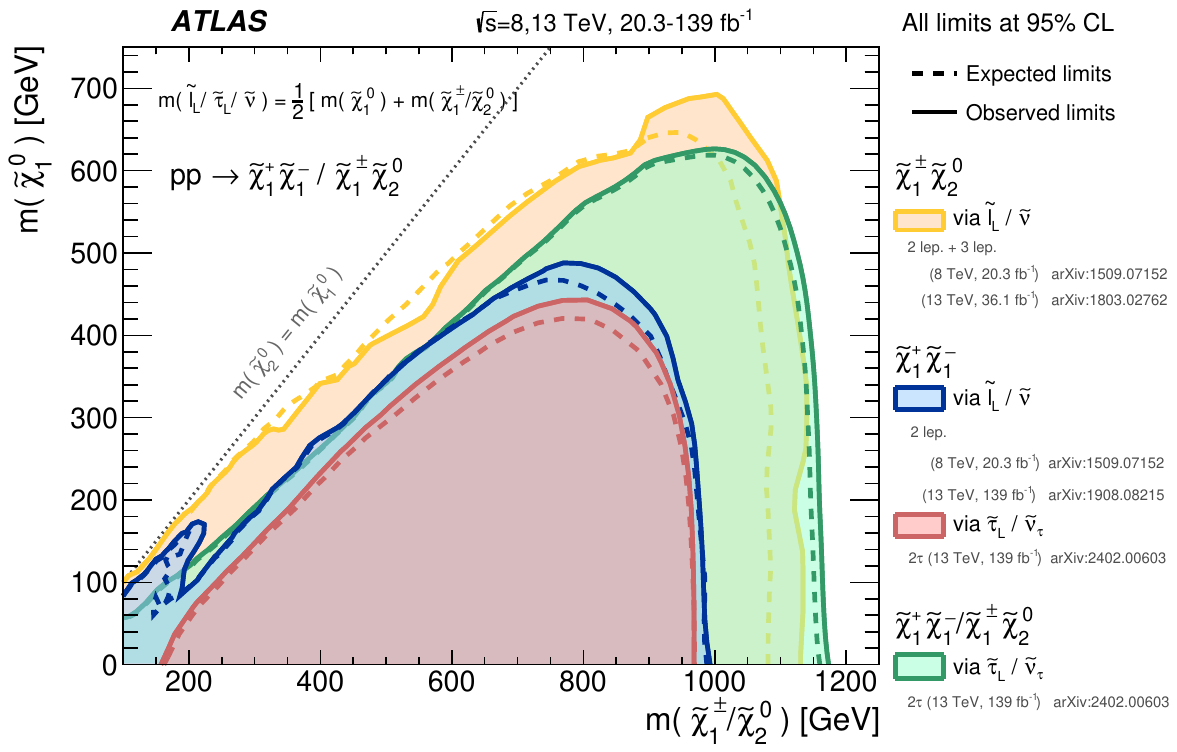}
\caption{The exclusion limits on $\chinoonep\chinoonem$ and $\chinoonepm\ninotwo$ production with $\slepton$-mediated decays, as a function of the $\chinoonepm$, $\ninotwo$ and $\ninoone$ masses. The production cross-section is for pure-wino $\chinoonep\chinoonem$ and $\chinoonepm\ninotwo$. Where more than one reference is given for each line, the total exclusion limit from multiple analyses is shown.}
\label{fig:WinoPair_viaSlep}
\end{figure}

Sleptons may also be produced as intermediate states in chargino and neutralino decay chains.
The production of chargino pairs or chargino--neutralino pairs decaying via sleptons results in two-lepton or three-lepton final states, respectively, along with $\met$.
The exclusion limits in these models set by the searches in  Refs.~\cite{SUSY-2018-32,SUSY-2016-24,SUSY-2023-03} are shown in Figure~\ref{fig:WinoPair_viaSlep}, where chargino production is excluded for chargino masses up to $1\,\TeV$ and chargino--neutralino production is excluded for masses up to $1.15\,\TeV$.

\begin{figure}[t]
\centering
\includegraphics[width=0.6\textwidth]{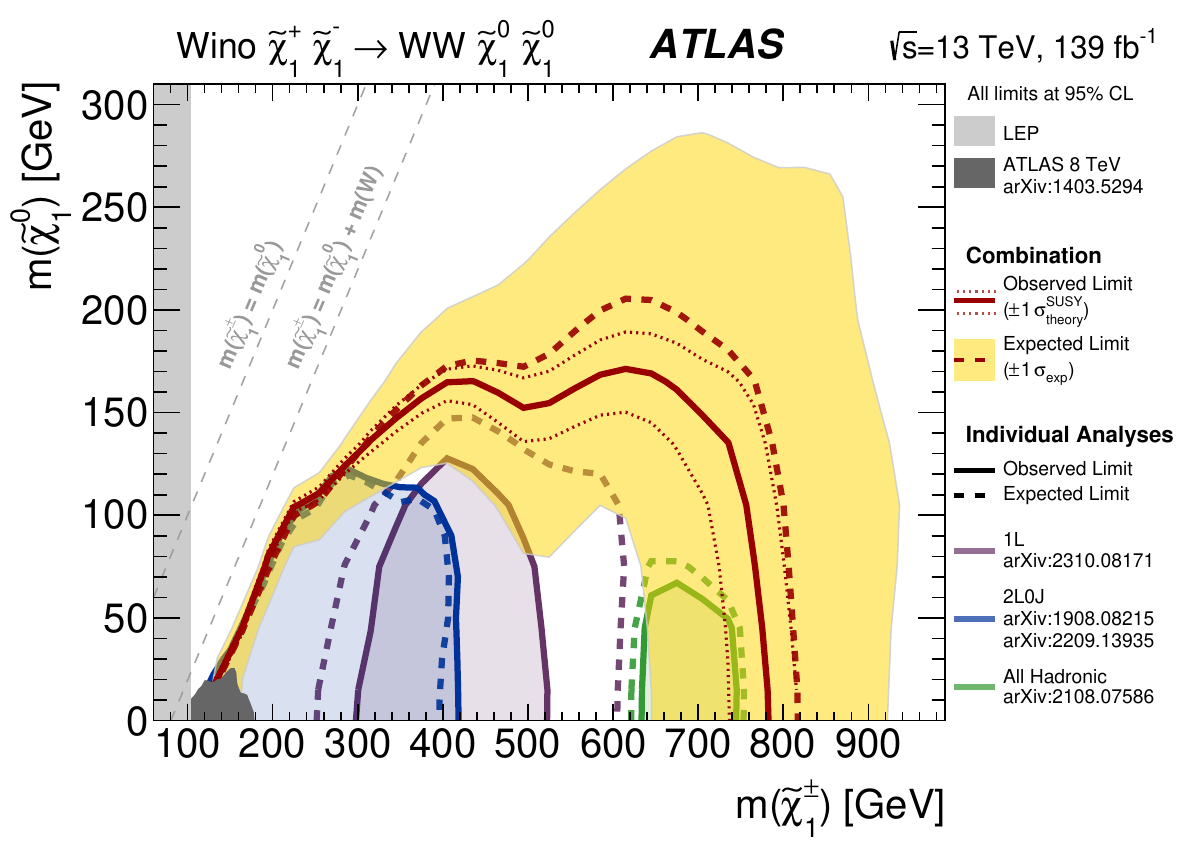}
\caption{The exclusion limits on $\chinoonep\chinoonem$ production with $W$-boson-mediated decays, as a function of the $\chinoonepm$ and $\ninoone$ masses. The production cross-section is for pure-wino $\chinoonep\chinoonem$. The labels with L and J denote the number of leptons and jets in the search, respectively. Where more than one reference is given for each line, the total exclusion limit from multiple analyses is shown.}
\label{fig:WinoPair_viaWbosons}
\end{figure}

Charginos and neutralinos will decay via SM $W$, $Z$ or $h$ bosons if decays via sleptons are not kinematically allowed.
The decays from the SM bosons include a rich array of hadronic and leptonic final states, allowing searches to be performed using jets only, or one, two or three leptons in the final state.
The same two-lepton search described above for slepton production is also used to search for the production of low-mass chargino pairs decaying via $W$ bosons~\cite{SUSY-2018-32}, with further searches improving sensitivity to the $\Delta m\!\left(\chinoonepm,\ninoone\right)\sim \mass{W}$ scenarios~\cite{SUSY-2019-02}.
The all-hadronic search~\cite{SUSY-2018-41} makes use of the large hadronic branching fraction of SM boson decays to target very high-mass chargino production with large $\Delta m\!\left(\chinoonepm,\ninoone\right)$.
The large mass splittings lead to boosted topologies, motivating the use of large-radius jets and boson-tagging algorithms to identify potential signal from boosted SM boson decays.
Finally, a one-lepton, two-jet final state search targets the intermediate chargino mass scenarios~\cite{SUSY-2023-01}, which are not covered by the all-hadronic and two-lepton searches.
No significant excesses were observed in the searches for chargino pair production, and a statistical combination of the three channels is performed in Ref.~\cite{SUSY-2020-05} %
and shown in Figure~\ref{fig:WinoPair_viaWbosons} for pure-wino charginos.
Chargino masses up to $780\,\GeV$ are excluded for massless $\ninoone$, and $\ninoone$ masses up to $170\,\GeV$ are excluded for $\chinoonepm$ masses of ${\sim}400{-}600\,\GeV$.

\begin{figure}[t]
\centering
\subfigure[]{\includegraphics[width=0.45\textwidth]{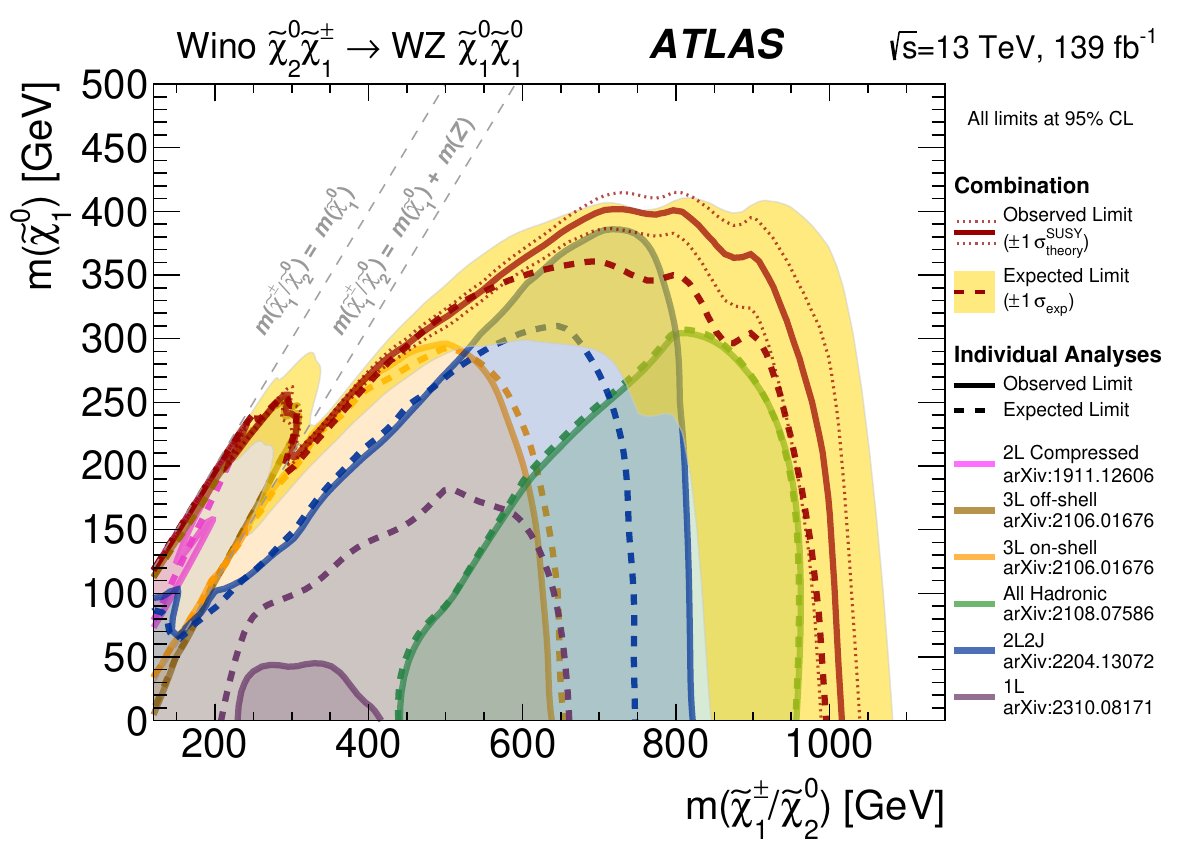}\label{fig:WinoPair_viaWZ}}
\subfigure[]{\includegraphics[width=0.45\textwidth]{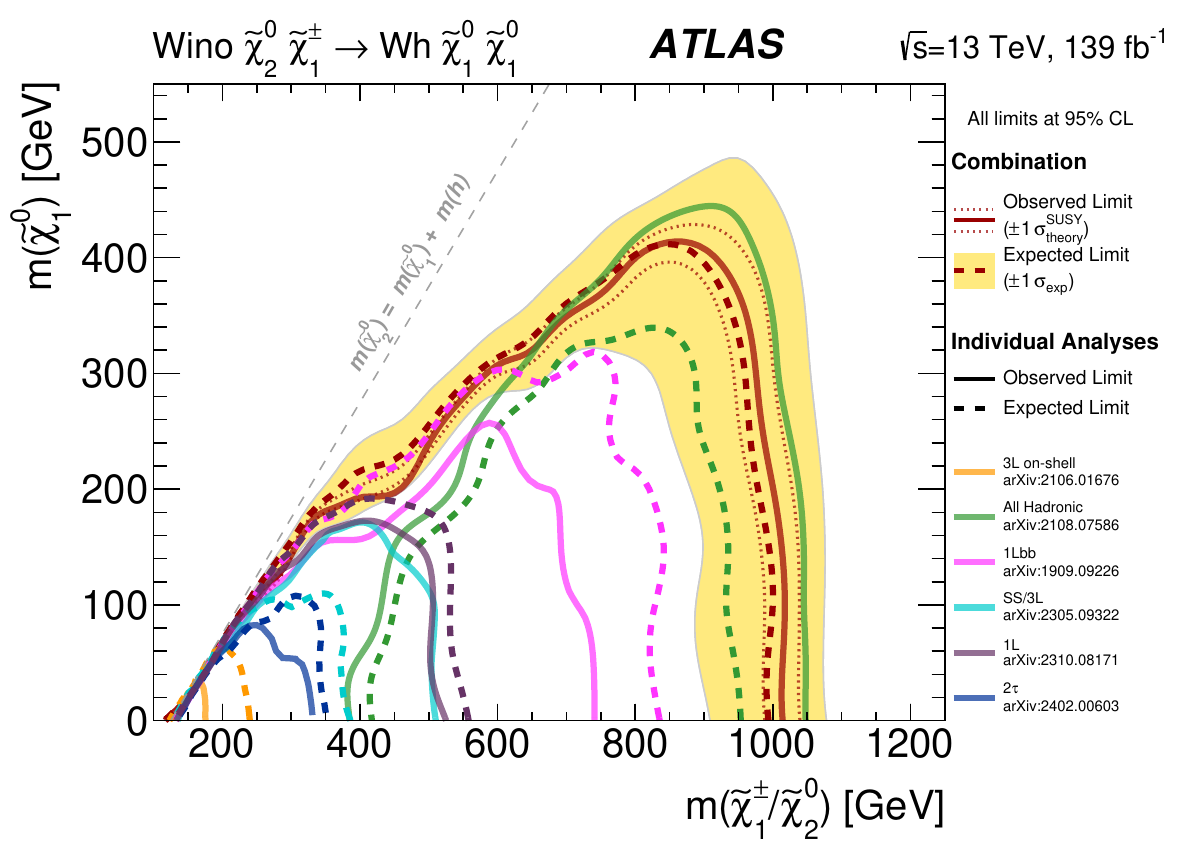}\label{fig:WinoPair_viaWh}}
\caption{The exclusion limits on $\chinoonepm\ninotwo$ production with $\chinoonepm\rightarrow\ninoone W^\pm$ and (a) $\ninotwo\rightarrow \ninoone Z$  or (b) $\ninotwo\rightarrow \ninoone h$, as a function of the $\chinoonepm$, $\ninotwo$ and $\ninoone$ masses. The production cross-section is for pure-wino $\chinoonepm$ and $\ninotwo$. The labels with L and J denote the number of leptons and jets in the search, respectively.}
\label{fig:WinoPair_viaWZh}
\end{figure}

The all-hadronic channel is also powerful in the search for chargino--neutralino production and dominates the sensitivity to high-mass, large-$\Delta m\!\left(\chinoonepm/\ninotwo,\ninoone\right)$ scenarios with decays via SM bosons, as seen in Figures~\ref{fig:WinoPair_viaWZ} and \ref{fig:WinoPair_viaWh}.
A two-lepton, two-jet search~\cite{SUSY-2018-05} selecting $Z\rightarrow \ell\ell$ and $W\rightarrow qq$ events with large $\met$ is sensitive to moderately high-mass $\chinoonepm\ninotwo$ scenarios with decays via $W$ and $Z$ bosons.
A few of the two-lepton, two-jet signal regions with closely spaced jets observed 2$\sigma$ deficits, resulting in a stronger observed limit than expected.
The sensitivity to moderately low-mass $\chinoonepm\ninotwo\rightarrow WZ\ninoone \ninoone$ scenarios is dominated by the three-lepton channel~\cite{SUSY-2019-09}, which selects for decays of on- and off-shell $W/Z$ bosons, with signal regions binned in OS~SF lepton invariant mass, $\met$, $\mt$, and hadronic activity.
Finally, the two-lepton compressed analysis~\cite{SUSY-2019-02} is sensitive to $\chinoonepm\ninotwo$ $WZ$ scenarios with the smallest mass splittings.

When the $\ninotwo$ decays via a Higgs boson, the one-lepton plus two-$b$-jet channel~\cite{SUSY-2019-08} is almost as powerful as the all-hadronic channel, and has unique sensitivity to moderate-mass scenarios.
Here, pairs of $b$-tagged jets consistent with a $h\rightarrow b \bar{b}$ decay, as well as high $\met$, $\mt$ and $\mct$, are used to suppress SM processes.
Leptonic final states are also used to search for these scenarios, but the sensitivity is limited to low masses due to the low branching fraction of the Higgs boson decay to leptons (possibly via SM bosons)~\cite{SUSY-2019-09,ATLAS:2023lfr,SUSY-2023-03}.
A statistical combination of all the channels targeting chargino--neutralino production is performed in Ref.~\cite{SUSY-2020-05} and shown in Figures~\ref{fig:WinoPair_viaWZ} and \ref{fig:WinoPair_viaWh}.
Pure-wino chargino--neutralino production is excluded up to $\chinoonepm/\ninotwo$ masses of ${\sim}1000\,\GeV$ for massless $\ninoone$, and up to $\ninoone$ masses of ${\sim}400\,\GeV$ for $\chinoonepm/\ninotwo$ masses of ${\sim}800\,\GeV$, independently of the $\ninotwo$ decay into a $Z$ or a Higgs boson.
For decays solely through $W$ and $Z$ bosons, $\chinoonepm/\ninotwo$ masses up to $300\,\GeV$ are excluded for $\Delta m(\chinoonepm/\ninotwo,\ninoone) \sim \mass{Z}$, decreasing to $240\,\GeV$ for $\Delta m(\chinoonepm/\ninotwo,\ninoone) =10\,\GeV$.

\begin{figure}[t]
\centering
\includegraphics[width=0.6\textwidth]{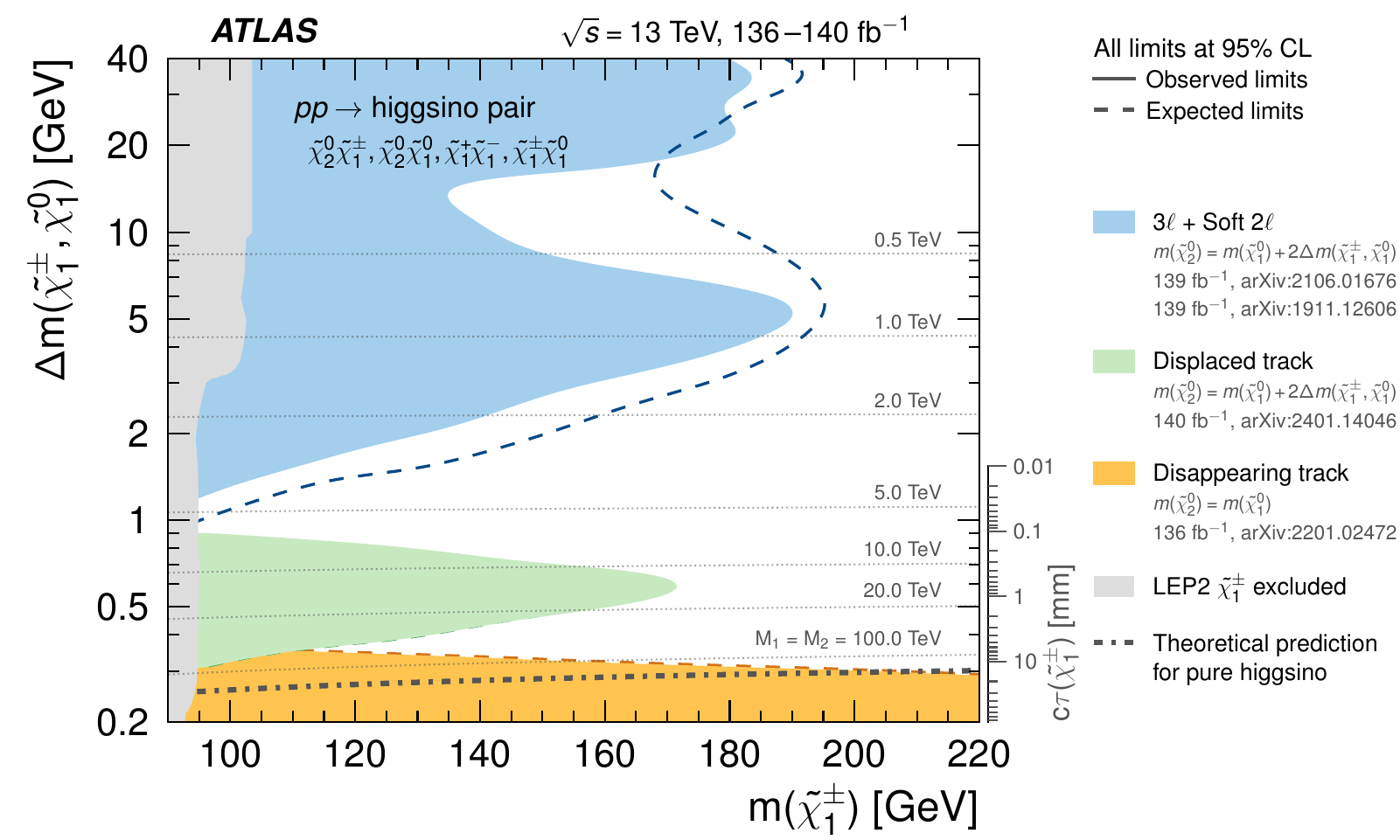}
\caption{Exclusion limits for higgsino pair production $\chinoonep\chinoonem$, $\chinoonepm\ninoone$, $\chinoonepm\ninotwo$, and $\ninoone\ninotwo$ with off-shell SM-boson-mediated decays to the lightest neutralino, $\ninoone$, as a function of the $\chinoonepm$ and $\ninoone$ masses. The production cross-section is for pure higgsinos. Where more than one reference is given for each line, the total exclusion limit from multiple analyses is shown. The LEP2 $\chinoonepm$ exclusion is taken from \protect Ref.~\cite{LEPSUSYWG:02-04.1}.}
\label{fig:HiggsinoPair_Compressed}
\end{figure}

The higgsino mass parameter $\mu$ enters into the MSSM expression for the light Higgs boson's mass at tree level~\cite{Heinemeyer:2004ms}. Therefore, a relatively low $\mu$  (of the order of a few hundred \GeV\ at most) is one of the firm predictions of naturalness arguments applied to SUSY~\cite{Baer:2012uy}. The search for pure-higgsino chargino--neutralino production is driven by the two-lepton compressed~\cite{SUSY-2018-16} and three-lepton analyses~\cite{SUSY-2019-09} described above, as well as the disappearing-track analysis\footnote{To obtain long-lived higgsinos, unnaturally high values of $M_1, M_2$ need to be invoked: the disappearing track analysis was initially designed to target scenarios with a wino-like LSP multiplet of electroweakinos.}~\cite{SUSY-2018-19} (see Section~\ref{sec:LLP}).
The off-shell regions of the three-lepton analysis provide sensitivity to the larger mass splittings between the produced higgsinos and the LSP, $\Delta m(\chinoonepm,\ninoone)>15\,\GeV$, while the more compressed analysis covers the smaller mass splittings down to about $1\,\GeV$.
A statistical combination of the two-lepton compressed analysis and three-lepton analysis is performed in Ref.~\cite{SUSY-2019-09}, excluding pure-higgsino production of charginos with masses up to $180\,\GeV$ ($190\,\GeV$) for $\Delta m(\chinoonepm,\ninoone)\sim 30\,\GeV$ ($ 5\,\GeV$). This decreases to $135\,\GeV$ for the intermediate mass splittings, where the combination has the most impact compared to the individual analyses.
For $\Delta m(\chinoonepm,\ninoone) <  1\,\GeV$, the pair-produced higgsinos start to have a flight path of a fraction of a mm, and their decay products are mildly displaced with respect to the primary vertex.
An innovative analysis, exploiting the presence of a low-\pt, isolated, good-quality track with only loose requirements on its impact parameter targets the prompt decay of the produced charginos and neutralinos into $\ninoone$, achieving sensitivity up to chargino masses of about $170\,\GeV$ for mass splittings $0.3\,\GeV < \Delta m(\chinoonepm,\ninoone)<0.9\,\GeV$~\cite{ATLAS:2024umc}. For the smallest mass splittings, the chargino is long-lived and is targeted by the disappearing-track analysis, where chargino masses below $210\,\GeV$ are excluded for $\Delta m(\chinoonepm,\ninoone) \lesssim 0.3\,\GeV$. All these results are summarised in Figure~\ref{fig:HiggsinoPair_Compressed}.

\begin{figure}[t]
\centering
\includegraphics[width=0.6\textwidth]{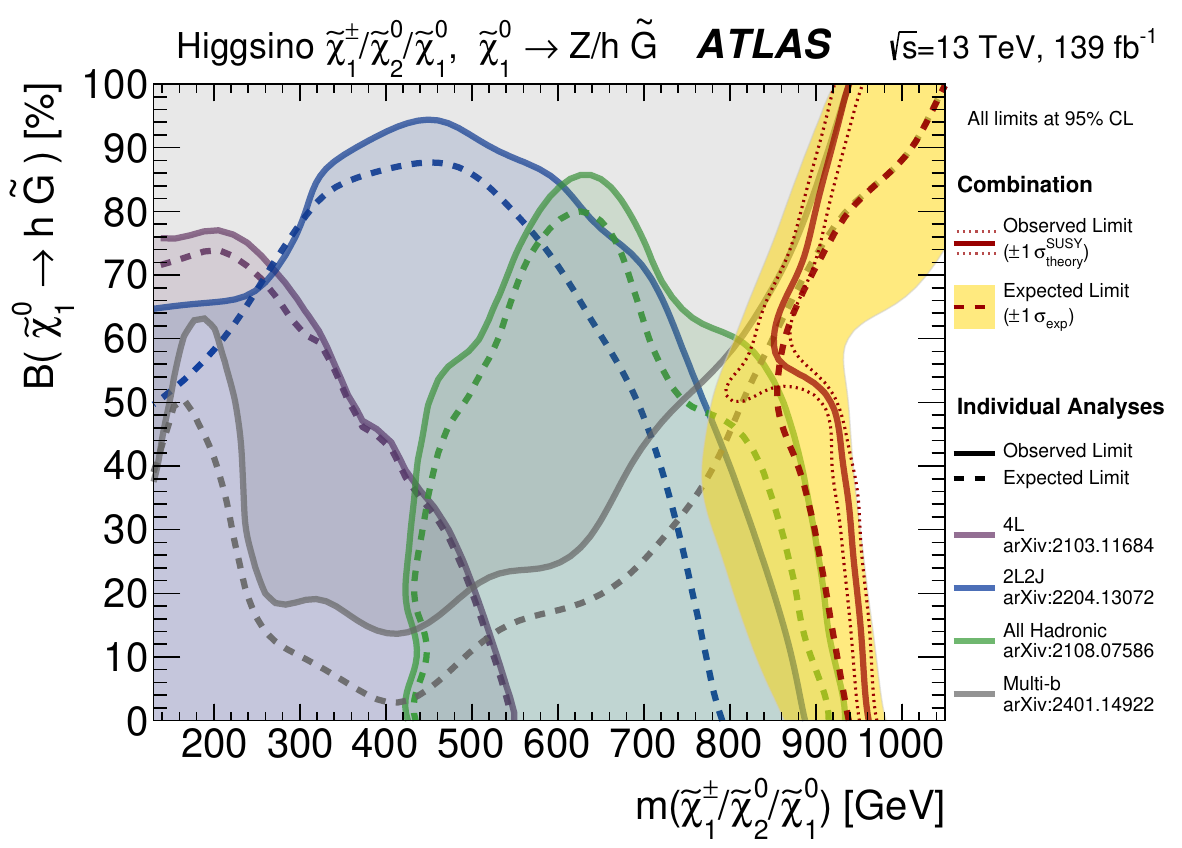}
\caption{Exclusion limits in a General Gauge Mediation model from $13\,\TeV$ data. The model assumes a pure-higgsino NLSP that promptly decays into either a $Z$ boson and gravitino or a Higgs boson and gravitino. The limits are displayed as a function of the mass of the nearly mass-degenerate higgsino multiplet and the branching fraction of the lightest higgsino into a Higgs boson and gravitino.}
\label{fig:HiggsinoPair_GGM}
\end{figure}

Higgsino GGM scenarios are targeted by numerous ATLAS SUSY analyses selecting for either the leptonic or hadronic decays of the $Z$ boson, or Higgs boson decays to $b$-quarks.
A statistical combination of the channels targeting higgsino GGM models is performed in Ref.~\cite{SUSY-2020-05} and shown in Figure~\ref{fig:HiggsinoPair_GGM}.
For high $\ninoone \rightarrow Z \gravino$ branching fractions, the four-lepton search~\cite{SUSY-2018-02} and all-hadronic search~\cite{SUSY-2018-41} both select for two $Z$ candidates and high $\met$, and dominate the sensitivity to low and high higgsino masses, respectively.
The two-lepton, two-jet channel~\cite{SUSY-2018-05} spans the intermediate higgsino mass range and extends to the lowest $\ninoone \rightarrow Z \gravino$ branching fraction in the searches involving a $Z$ boson.
The multi-$b$-jet search~\cite{SUSY-2020-16} %
and two-$\gamma$ two-$b$-jet search~\cite{ATLAS-CONF-2023-009} target GGM scenarios with high $\ninoone \rightarrow h \gravino$ branching fractions.
The multi-$b$-jet channel makes use of the common Higgs boson decay into $b\bar{b}$, with signal regions binned in $\met$ and $\meff$ for low-mass scenarios, or selecting on a BDT parameterized in higgsino mass for high-mass scenarios.
The $\ninoone$ branching fraction into either $Z \gravino$ or $h\gravino$ is fully covered by the individual searches, with higgsino masses up to $960\,\GeV$ excluded for the two extreme branching fraction scenarios.
The sensitivity drops to masses of $850\,\GeV$ for the mixed branching fraction scenarios, where the combination has the most impact.

\section{R-parity-violating decays}
\label{sec:RPV}

%

The MSSM superpotential includes terms that violate the conservation of baryon number and lepton number,
\begin{equation*}
\frac{1}{2}\lambda_{ijk} L_iL_j\bar{E}_k +   \lambda_{ijk}' L_iQ_j\bar{D}_k    + \frac{1}{2}\lambda_{ijk}'' \bar{U_i}\bar{D_j}\bar{D_k}   +\kappa_iL_iH_2,
\end{equation*}
where $L_i$ and $Q_i$ are the lepton and quark $\mathrm{SU}(2)$-doublet superfields, respectively, and $\bar{E_i}$, $\bar{U_i}$ and $\bar{D_i}$ are the corresponding singlet superfields.
The fermion generations are denoted by the indices $i$, $j$ and $k$, while the Higgs field that couples to up-type quarks is represented by the Higgs $\mathrm{SU(2)}$-doublet superfield $H_2$.
The $\lambda_{ijk}$, $\lambda_{ijk}'$ and $\lambda_{ijk}''$ parameters are three sets of new Yukawa couplings, while the $\kappa_i$ parameters have dimensions of mass. Consistency with limits on barion and lepton number conservation impose severe constraints on many of these couplings, nevertheless leaving the possibility that the least constrained ones (typically connected with the third fermion generations) are sizeable.
In this section, RPV couplings are assumed to be large enough to allow decays to be prompt.
In the case of very small RPV couplings, SUSY particles could be long-lived, leading to rather different experimental signatures which are discussed in Section~\ref{sec:LLP}.

\begin{figure}[t]
\centering
\includegraphics[width=0.6\textwidth]{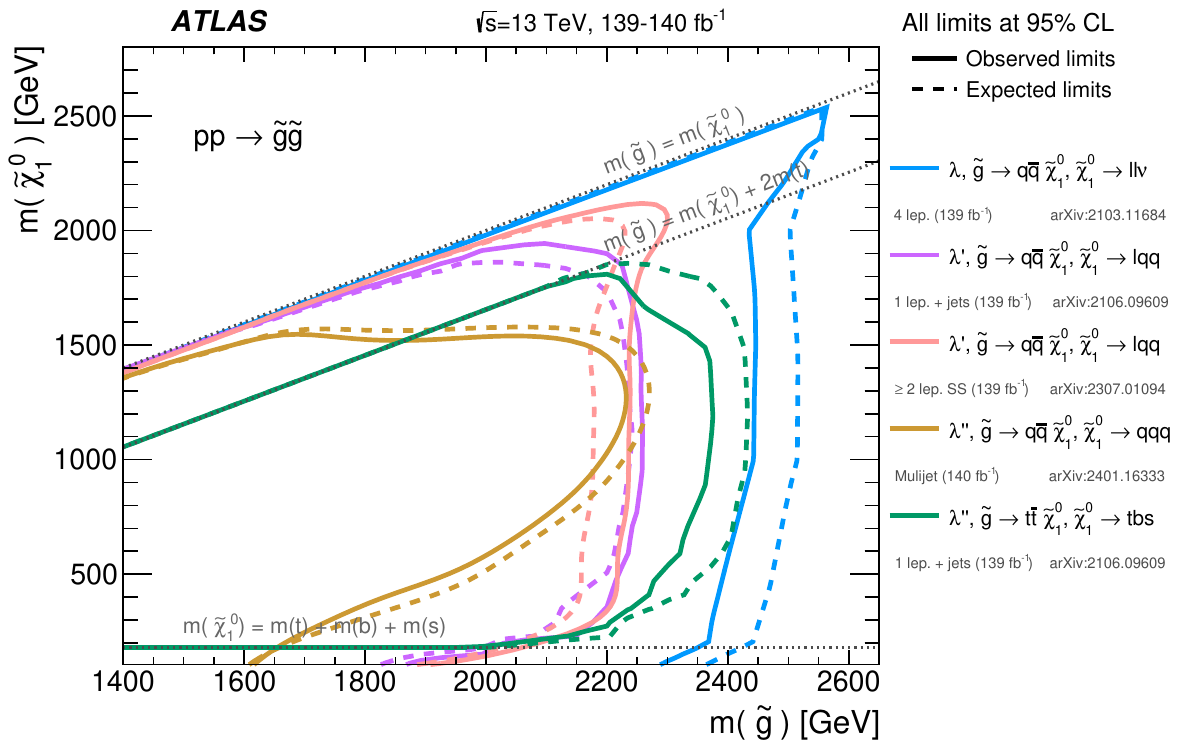}
\caption{Exclusion limits at 95\% CL based on $13\,\TeV$ data in the $(\gluino, \ninoone)$ mass plane for different simplified models featuring the decay of the gluino to the lightest supersymmetric particle ($\ninoone$) which in turn decays via R-parity-violating couplings to SM particles. For each line, the decay mode is reported in the legend and it is assumed to proceed with 100\% branching fraction. Some limits depend on additional assumptions, as described in the references cited in the plot.}
\label{fig:LLP_Gluino_RPV}
\end{figure}

The nine $\lambda_{ijk}$ couplings violate lepton number conservation via interactions between the three generations of leptons and sleptons, allowing a $\ninoone$ LSP to decay via virtual sleptons into $\ell^{\pm}_k \ell^{\mp}_{i/j} \nu_{j/i}$.
Pair production of a promptly decaying NLSP is assumed, potentially producing four or more charged leptons from the two $\ninoone$ decays in the event.
These scenarios were explored with the four-lepton final state in Ref.~\cite{SUSY-2018-02}, selecting electrons, muons and hadronically decaying $\tau$-leptons for sensitivity to the $\lambda_{12k}$ and $\lambda_{i33}$ couplings, where $i,k = 1,2$.
Since the electron and muon reconstruction efficiencies at the ATLAS experiment are similarly high, the sensitivities to $\lambda_{121}$ and $\lambda_{122}$ scenarios are nearly identical~\cite{SUSY-2013-13} and can be considered simply as $\lambda_{12k}$ (and similarly for $\lambda_{i33}$).
The $\lambda_{12k}\neq 0$ couplings are targeted using electrons and muons only, while sensitivity to the $\lambda_{i33}\neq 0$ couplings is increased by including up to two hadronically decaying $\tau$-leptons among the four leptons.
For each coupling scenario, three different NLSP production processes are considered: wino pair production ($\chinoonepm/\ninotwo/\chinoonep\chinoonem$), slepton pair production ($\slepton\slepton/\snu\snu/\slepton\snu$), or gluino pair production ($\gluino\gluino$).
The strongest limits are set on $\gluino\gluino$ production with $\lambda_{12k}\neq 0$ as shown in Figure~\ref{fig:LLP_Gluino_RPV}, where gluino masses up to $2.5\,\TeV$ are excluded. This mass limit decreases to $1.9\,\TeV$ for $\lambda_{i33}\neq 0$ scenarios (not shown in Figure~\ref{fig:LLP_Gluino_RPV}), due to the lower reconstruction efficiency for hadronic $\tau$-lepton decays.
Lower mass exclusion limits of $1.6\,\TeV$ and $1.2\,\TeV$ ($1.1\,\TeV$ and $0.9\,\TeV$) are set on $\lambda_{12k}\neq 0$ ($\lambda_{i33}\neq 0$) scenarios of wino pair production and slepton pair production, respectively, due to the lower cross-sections.
The sensitivity to other $\lambda_{ijk}$ scenarios is expected to fall between the two extremes of the $\lambda_{12k}$ and $\lambda_{i33}$ scenarios studied.

The 27 $\lambda_{ijk}'$ couplings violate lepton and baryon number conservation with interactions between leptons, quarks, and squarks, allowing a $\ninoone$ LSP to now decay via virtual squarks into $\ell^{\pm}_k q_i \bar{q}_j / \nu_k q_i q_j$, where $q_i$ and $\bar{q}_j$ may be up-type or down-type quarks as needed to conserve charge.
The lepton from one or each $\ninoone$ decay, along with several jets, is used to probe models of gluino pair production.
The one-lepton analysis~\cite{SUSY-2019-04} divides the search regions according to the number of jets and number of $b$-tagged jets, as well as the charge of the lepton.
Increasing thresholds are set on the $\pt$ of the counted jets to obtain sensitivity to a wide range of scenarios.
A second search selects two leptons with same-sign charges~\cite{ATLAS:2023afl}, multiple jets and high $\meff$, benefiting from a lower SM background than in the one-lepton search.
Figure~\ref{fig:LLP_Gluino_RPV} shows that the two searches set similar mass limits in the gluino production models, excluding gluinos with masses below ${\sim}2.2\,\TeV$.
The two-lepton search has less reliance on reconstructing the jet from the gluino decay ($\gluino \rightarrow qq\ninoone$) and has a slight advantage if $m(\gluino)-m(\ninoone)$ is small.

Finally, the nine $\lambda_{ijk}''$ couplings violate baryon number conservation with interactions between quarks and squarks, allowing a $\ninoone$ LSP to now decay via virtual squarks into $q_k q_i q_j$, where $q_k$ is an up-type quark and $q_{i/j}$ are down-type quarks. The one-lepton analysis already mentioned~\cite{SUSY-2019-04} also sets (for the first time) limits on higgsino pair production followed by their decay via $\lambda''$ couplings into third-generation quarks.
The multijet search~\cite{SUSY-2019-24} selects events with a large number of high-$\pt$ jets, allowing the presence of $b$-tagged jets to increase the sensitivity to $b$-quarks from the RPV $\ninoone$ decay.
Again, gluinos with masses below ${\sim}2.2\,\TeV$ are excluded in Figure~\ref{fig:LLP_Gluino_RPV}, but the sensitivity drops off rapidly for high and low $\ninoone$ masses where the jets are either too low in momentum or too collimated to be reconstructed.
The analysis uses an approach similar to that of the one-lepton search described above~\cite{SUSY-2019-04}, which itself is also sensitive to the  $\lambda_{ijk}''\neq 0$ scenarios when leptons are produced in top quark decays.
The one-lepton signal regions with many $b$-tagged jets have good sensitivity to the $\gluino\rightarrow tt\ninoone$, $\ninoone\rightarrow tbs$ scenarios, excluding gluinos with masses below ${\sim}2.35\,\TeV$ as shown in Figure~\ref{fig:LLP_Gluino_RPV}.
A second approach in Ref.~\cite{SUSY-2019-24} uses a neural network with an architecture based on transformers to group jets together in order to help deal with the large combinatorial problem caused by the large number of jets in the final state, and search for a mass resonance from the two gluinos decaying directly via the $\lambda_{ijk}''$ coupling, $\gluino\rightarrow qqq$. In this case, gluinos with masses below ${\sim}1.8\,\TeV$ are excluded (not shown in Figure~\ref{fig:LLP_Gluino_RPV}).

Less simple RPV SUSY scenarios have also been explored by the ATLAS Collaboration.
For example, an MSSM scenario explored in Ref.~\cite{SUSY-2018-36} adds $B-L$ symmetry breaking with right-handed sneutrinos that is small enough to satisfy proton decay constraints, and also adds $L$ symmetry breaking at tree level.
The latter gives rise to charginos and neutralinos that decay into a SM boson and a lepton, e.g.\ $\chinoonepm\rightarrow Z \ell$, where a fully leptonic decay of the $Z$ boson allows the chargino mass resonance to be fully reconstructed.
The invariant mass of the three leptons is binned to achieve good sensitivity to a wide range of these $B-L$ SUSY scenarios, and the presence of additional leptons can be used to identify the presence of SM boson decays from the second produced sparticle.
With no significant excess seen in the data, chargino masses up to $1.1\,\TeV$ are excluded for $\mathcal{B}(\chinoonepm\rightarrow Z \mu)=100\%$, with lower mass exclusion limits obtained with different assumptions.

No direct limits have yet been placed on the size of the RPV couplings. However, if the coupling is not particularly small, the prompt sparticle decay is allowed and its effect is included in the limits set here.
When the RPV coupling becomes very small, the sparticle becomes long-lived and experimental signatures are rather different -- these are the topic of Section~\ref{sec:LLP}.


%
\section{Long-lived supersymmetric particles}
\label{sec:LLP}
%

As is the case for SM particles, SUSY particles produced at the LHC may be long-lived and travel a significant distance before decaying.
Longer lifetimes may be due to weak couplings to their decay products, decays through heavy mediator particles, or small mass differences between the particle and the decay products.
The experimental signatures of long-lived new particles can be unconventional and depend on where and how the SUSY particle decays.
A long-lived SUSY particle could travel from the ATLAS interaction point, through the inner-detector tracking system and then the calorimeters, and even through the muon spectrometer (or decay at any point along the way), leaving different signatures along its path depending on its properties.
A charged particle decaying in the inner detector to a nearly degenerate stable neutral particle will manifest itself as a disappearing (or kinked) track, while a neutral particle's decay in the inner detector to charged and neutral particles would appear as tracks pointing back to a displaced vertex.
The calorimeters may also be used for long-lived particle signatures, with photons that do not point back to the original interaction point (\enquote{non-pointing} photons), or strongly interacting long-lived particles that are stopped in the calorimeters by ionisation energy loss and nuclear scattering.
Finally, if a SUSY particle is charged and very long-lived, the experimental signature would be similar to that of a muon, albeit with high mass.
A neutral, weakly interacting long-lived new particle traversing the ATLAS detector would not be detected and would appear as a contribution to $\ptmiss$.

\begin{figure}[t]
\centering
\includegraphics[width=0.6\textwidth]{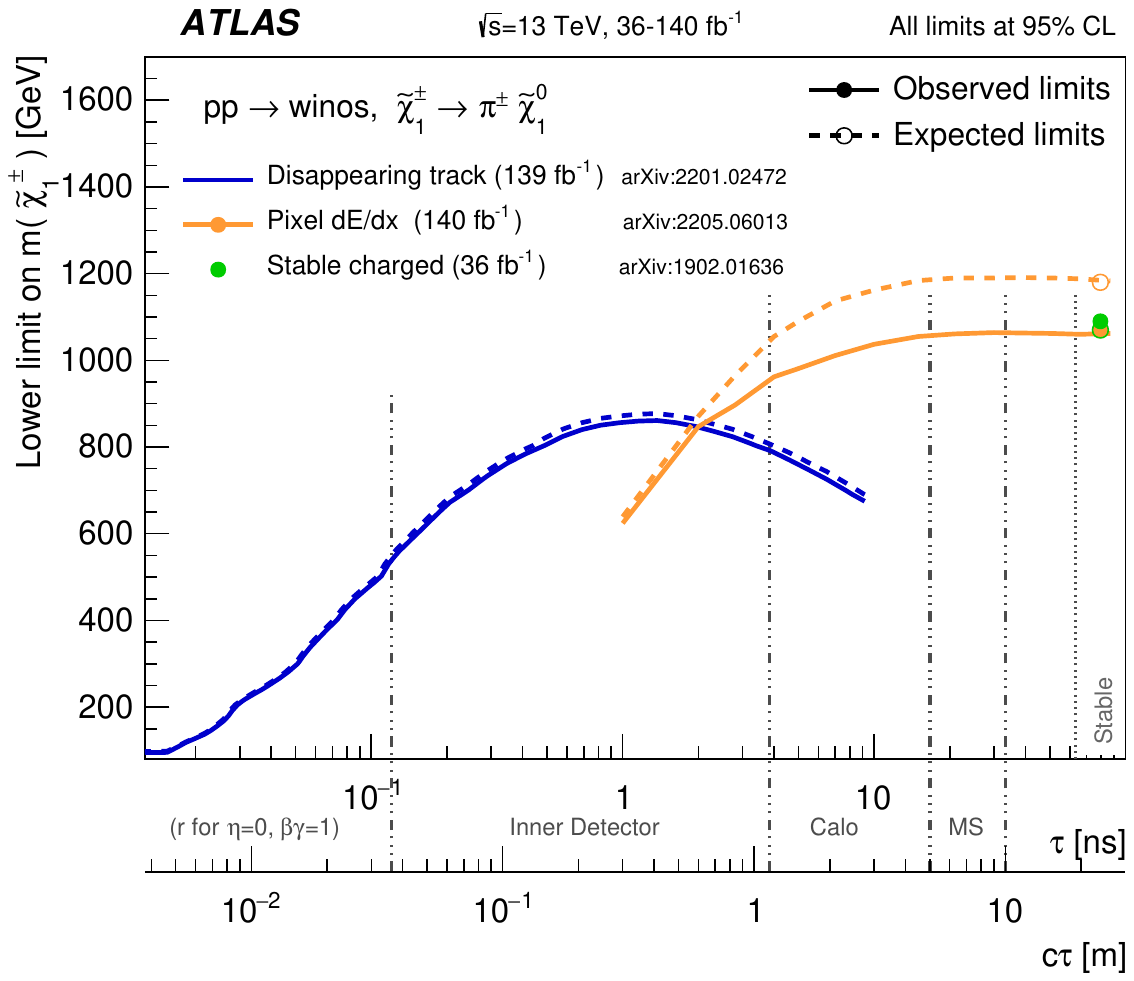}
\caption{Constraints on the  $\chinopm$ mass as a function of its lifetime for an AMSB model with $\tan\beta=5$ and $\mu>0$. The wino-like $\chinopm$ is pair-produced and decays into a wino-like $\ninoone$ and a very soft charged pion. The solid lines indicate the observed limits, while the dashed lines indicate the expected limits. The area below the curves is excluded. ``Calo'' and ``MS'' refer to the ATLAS calorimeter and muon spectrometer systems, respectively. The analyses also have sensitivity at lifetimes other than those shown, but only the limits at tested lifetimes are shown. The three markers at large lifetime represent results for which the particle is assumed to be stable. In this context, stable means escaping the detector. The exclusion contours from the pixel $\dif{E}/\dif{x}$ search are each extrapolated to the stable regime with a straight line.}
\label{fig:LLP_Chargino_AMSB}
\end{figure}

A long-lived charged SUSY particle, such as a slepton or chargino, may be produced directly in $pp$ collisions, or from prompt decays of other SUSY particles, such as a gluino.
If its lifetime is ${\sim}0.1{-}10$\,ns because it is nearly degenerate with the invisible LSP, the charged SUSY particle would leave a track in the inner detector until the point of decay, where that track then disappears (the emitted charged decay products typically have momenta which are too low for their tracks to be reconstructed efficiently by the standard algorithm, although there is enough activity to reconstruct the additional decay products with a dedicated technique~\cite{ATL-PHYS-PUB-2019-011}).
The backgrounds to such a signature include badly measured tracks, leptons undergoing large bremsstrahlung or scattering, and high-momentum charged hadrons interacting with material in the inner detector, and all are estimated directly from data using smeared tracks in control regions.
Additional objects in the events, such as large $\met$, are usually selected to improve sensitivity to particular SUSY scenarios.
The disappearing-track analysis~\cite{SUSY-2018-19} excludes pure-wino charginos with masses up to ${\sim}850\,\GeV$ for lifetimes of about $1\,$ns, as shown in Figure~\ref{fig:LLP_Chargino_AMSB}.
The mass sensitivity for lower lifetimes is limited by the requirement of a minimum number of hits on the different layers of the inner detector, while the veto on track extensions imposed to satisfy the \enquote{disappearing track} criterion limits the sensitivity at longer lifetimes.

If a charged particle has a longer lifetime, it may propagate beyond the inner detector before decaying.
The track it leaves would be rather distinctive, with high ionisation energy loss since the expected high mass of the SUSY particle dictates it must be moving slowly, with $\beta = v/c \ll 1$.
An analysis selecting tracks with large ionisation energy loss ($\dif{E}/\dif{x}$) in the pixel detector, instead of disappearing tracks, is used to search for longer-lived charged SUSY particles~\cite{SUSY-2018-42}. Careful corrections to the measured $\dif{E}/\dif{x}$ and reconstructed mass are made to account for the tracker's decreasing charge collection efficiency due to radiation damage as the integrated luminosity increases.
The sensitivity offered by the large-$\dif{E}/\dif{x}$ analysis is complementary to that of the disappearing-track analysis, covering the longer-lifetime scenarios.
Wino-like charginos with masses up to ${\sim}1050\,\GeV$ are excluded for lifetimes longer than $10\,$ns, as shown in Figure~\ref{fig:LLP_Chargino_AMSB}. The observed limits are weaker than the expected ones
because the number of observed events exceeds the background prediction, quantified by a global $Z$ significance of 3.3. This excess was investigated in a more recent analysis~\cite{ATLAS-CONF-2023-044} which considered measurements of time-of-flight to the calorimeter as well as $\dif{E}/\dif{x}$ in the pixel detector, and found compatibility with the background prediction.

\begin{figure}[t]
\centering
\includegraphics[width=0.6\textwidth]{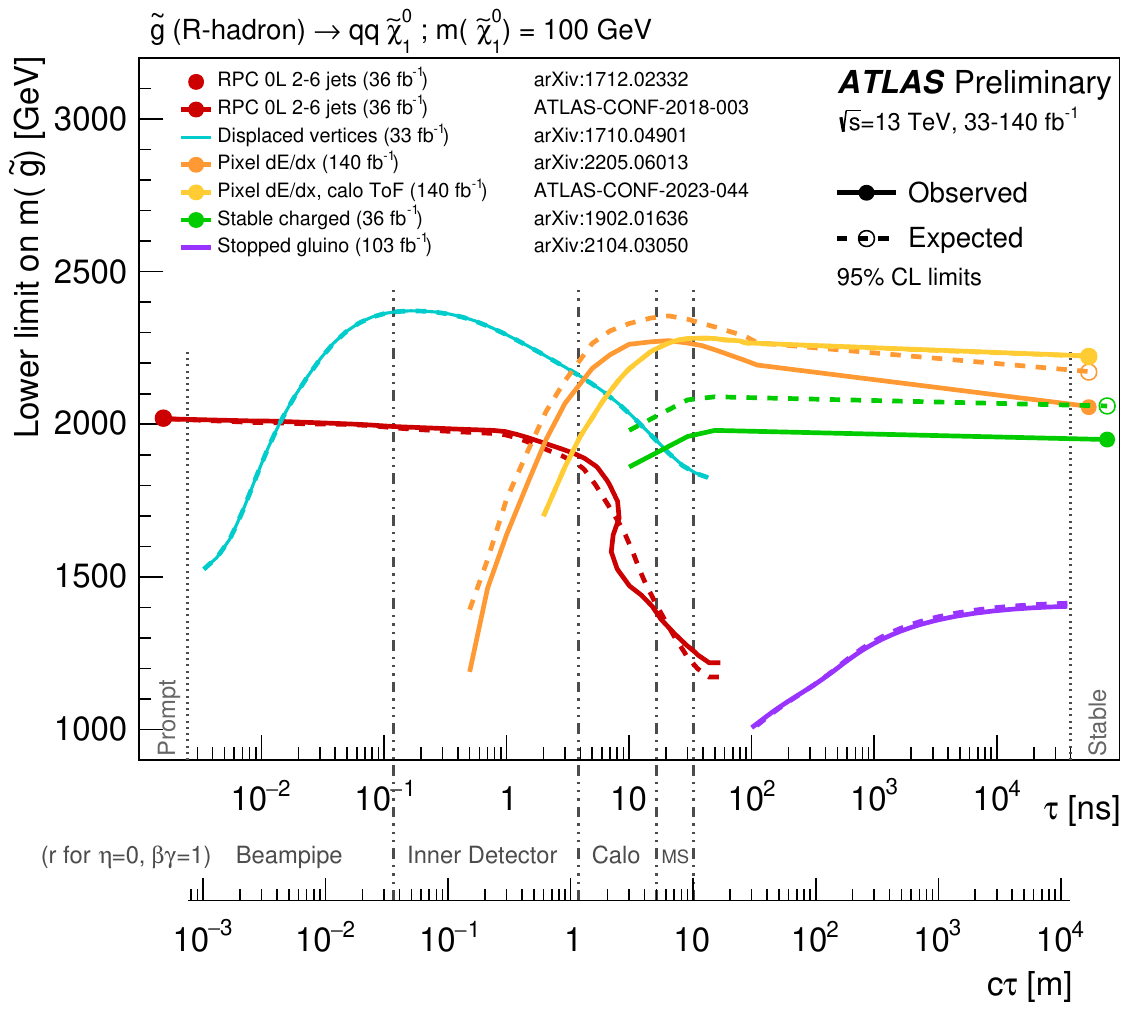}
\caption{Constraints on the gluino mass as a function of its lifetime lifetime for a split-supersymmetry model with the gluino R-hadron decaying into a gluon or light quarks and a neutralino with a mass of $100\,\GeV$. The solid lines indicate the observed limits, while the dashed lines indicate the expected limits. The regions below the curves are excluded. ``Calo'' and ``MS'' refer to the ATLAS calorimeter and muon spectrometer systems, respectively, while ``ToF'' refers to the use of time of flight measurements. The expected and observed lines coincide for the Pixel dE/dx, calo ToF search. The markers at large lifetime represent results for which the particle is assumed to be stable. In this context, stable means escaping the detector. }
\label{fig:LLP_G_Rhadron}
\end{figure}

An alternative approach, sensitive to neutral long-lived SUSY particles leaving no tell-tale track in the inner detector, is to reconstruct the long-lived particle's decay products.
Charged decay products leave tracks that can be traced back to a common displaced vertex, with those from SUSY decays easily distinguishable from SM decays by their large invariant mass.
A specialised track reconstruction algorithm optimised for tracks with large impact parameters was used to improve the efficiency of displaced-vertex reconstruction~\cite{ATL-PHYS-PUB-2017-014}.
Random track combinations and merged vertices mimicking a high-mass displaced vertex typically dominate the backgrounds.
A smaller component comes from hadronic interactions with detector material that are usually concentrated in regions with high matter density and are vetoed. However, those occurring in less dense regions can also appear as a high-mass displaced vertex.
The searches for long-lived SUSY particles using displaced vertices~\cite{SUSY-2018-13} have excluded gluino masses up to $2400\,\GeV$ for lifetimes of ${\sim}0.1\,$ns, as shown in Figure~\ref{fig:LLP_G_Rhadron}.
Prompt-decay SUSY analyses are typically more sensitive to gluino lifetimes below $0.01\,$ns, while the large-$\dif{E}/\dif{x}$ and stopped-gluino searches cover the longer lifetimes above ${\sim}5\,$ns.

A long-lived gluino may form an R-hadron with a SM vacuum quark, travel part way through the detector, and then stop in the calorimeter due to energy losses via ionisation and nuclear scattering.
It may decay at a much later time, from about $100\,$ns to one year later, so a search for hadronic activity in the absence of collisions, while empty LHC beam bunches pass through ATLAS, is used~\cite{SUSY-2018-15}.
The efficiency of R-hadron detection depends on the fraction that stop in the detector, the probability of R-hadron decay during empty bunches, and the efficiency of hadronic-activity reconstruction~\cite{ATL-PHYS-PUB-2019-019}.
The cosmic-ray muon background is taken from low-luminosity data-taking periods, while the beam-halo background is estimated from unpaired crossings.
Long-lived gluinos with masses up to $1400\,\GeV$ are excluded for lifetimes of $10^4{-}10^{12}\,$ns, as shown in Figure~\ref{fig:LLP_G_Rhadron}.

Reconstructed objects, such as electrons, muons, and photons, can also be used to search for long-lived SUSY particles.
In these cases, the leptons and photons should point back to the decay vertex of the SUSY particle rather than the interaction point, i.e.\ their tracks should be displaced.
The specialised tracking procedure in Ref.~\cite{ATL-PHYS-PUB-2017-014} reconstructs electrons and muons with large transverse impact parameters, $10\,\mathrm{mm}<|d_0|<300\,$mm.
The signature is very simple, requiring two large impact parameter ($|d_0|>3\,$mm) leptons as expected from the decay of a pair of sleptons~\cite{SUSY-2018-14}.
With no significant excess observed in the data, long-lived selectrons and smuons with masses up to about $700\,\GeV$ are excluded for lifetimes of about $0.1\,$ns (or, equivalently, co-NSLP slectron/smuon masses up to $820\,\GeV$), while long-lived staus with masses up to about $340\,\GeV$ are excluded for the same lifetime.
Another analysis, in Ref.~\cite{SUSY-2020-09}, covers shorter-lifetime slepton scenarios, in an attempt to cover any sensitivity gap between long-lived and promptly decaying sleptons: in this case, the analysis  uses standard tracking algorithms to select \enquote{micro-displaced} leptons with $0.1\,\mathrm{mm}<|d_0|<3\,$mm forming high-mass pairs.
With this approach, smuon masses up to $520\,\GeV$ are excluded for lifetimes of about $10^{-2}\,$ns.

Despite the lack of tracking information, the flight path of a photon can be traced back to a location other than the collision vertex by using the spatial measurement and timing capabilities of the ATLAS detector's liquid-argon EM calorimeter.
A photon from a long-lived particle decay will have delayed timing compared to promptly produced photons, and the longitudinal shape of the shower will point back to a displaced vertex instead of the collision vertex.
Two such non-pointing photons forming a high-mass displaced vertex are selected (assumed to be from the same decay) in Ref.~\cite{SUSY-2020-28}, and these non-pointing photons may also be interpreted as displaced electrons since only EM calorimeter information is used.
The background is estimated from data by exploiting the non-correlation in timing and pointing for photons from SM processes.
Long-lived neutralinos that decay into a $Z$ boson and gravitino and have lifetimes of $0.2\,$ns and masses below about $700\,\GeV$ are excluded.

In summary, an array of unconventional signatures in the ATLAS detector have been used to search for long-lived SUSY particles.
From the implementation of specialised tracking algorithms, to searching for activity in the absence of collisions, the effort extends beyond the boundaries of typical SUSY searches at ATLAS and sets stringent limits on the masses of SUSY particles with lifetimes spanning many orders of magnitude.


\section{Beyond simplified models}
\label{sec:pmssm}

%

Having summarised and discussed in Sections~\ref{sec:Strong} to~\ref{sec:LLP} the main exclusion limits obtained from ATLAS \RunTwo data when using simplified models, the obvious question is: how do these limits change when more realistic SUSY models are considered?

At the end of \RunOne, ATLAS produced two papers trying to answer this question using slightly different methodologies. Reference~\cite{SUSY-2014-08} used a general 19-parameter pMSSM~\cite{Berger:2008cq, Cahill-Rowley:2012ydr} and sampled the parameter space while assuming a flat prior within the chosen parameter range. The models were checked to see if they were excluded by any of the ATLAS analyses available at the time. Results were quoted in terms of the fraction of models excluded at a given sparticle mass. A further study was performed~\cite{SUSY-2015-12} by varying only the pMSSM parameters affecting the electroweak sector ($\tan \beta$, $M_1$, $M_2$, $\mu$, and $M_A$) and evaluating a global likelihood for models relevant for an explanation of the dark-matter relic density.

\begin{figure}[t]
\centering
\includegraphics[width=0.6\textwidth]{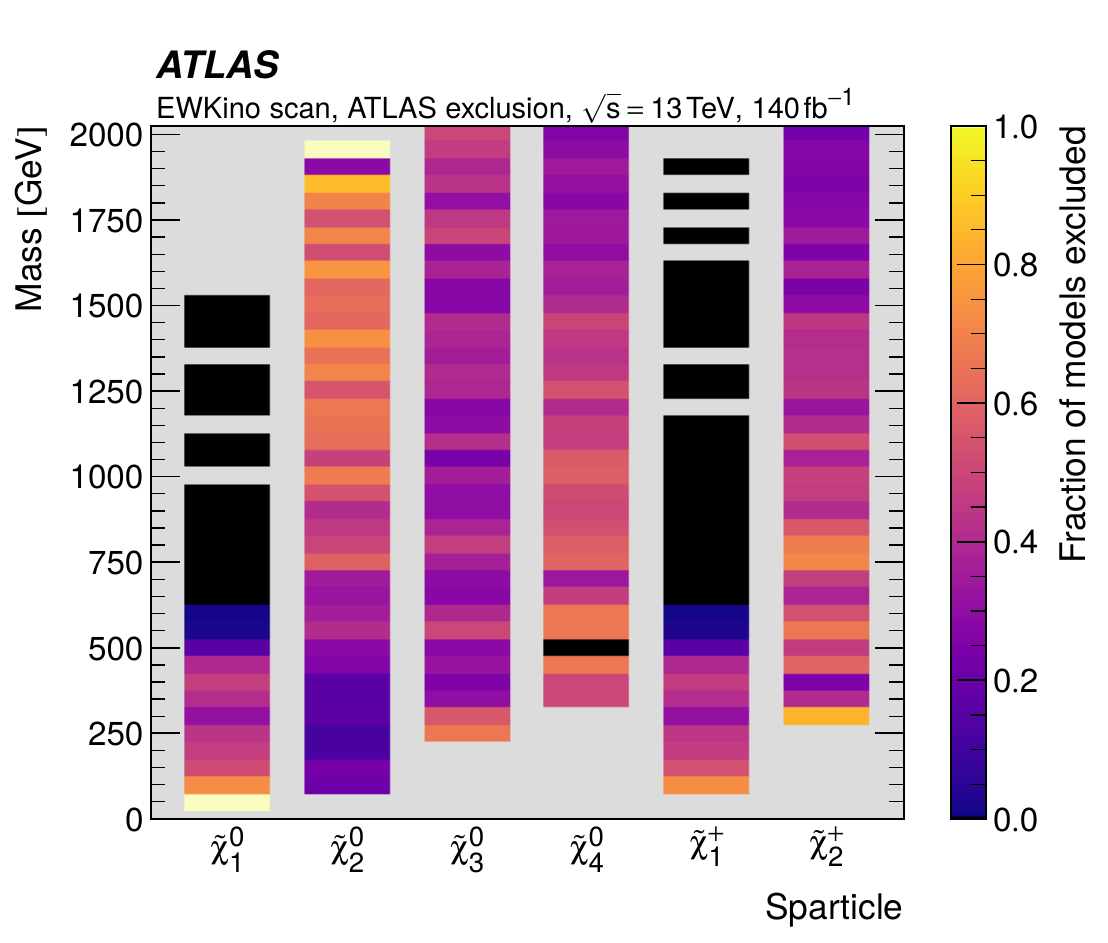}
\caption{Fraction of models passing external constraints but excluded by ATLAS, depending on the mass of the electroweakinos for the EWKino samples. Bins in grey indicate that no model passes external constraints, while for bins in cream (black) all models are (not) excluded.}
\label{fig:pmssm_frac_EWKino}
\end{figure}

A similar study was repeated at the end of Run 2~\cite{SUSY-2020-15}. The emphasis was again on the electroweak sector, with the gluino, squarks, and sleptons having very high masses. The five parameters mentioned above were varied (along with other parameters to satisfy constraints connected with the Higgs sector).  Two samplings were performed:
the \enquote{EWKino} sample started from 20\;\!000 models randomly sampled with uniform priors, while the \enquote{Bino-DM} sample started with 437\;\!500 models with a bino-like LSP that satisfy a requirement on the observed dark-matter relic density as an upper limit. The ability of several ATLAS \RunTwo analyses targeting electroweak production to exclude each of the models considered was assessed. A model is declared ``excluded by ATLAS'' if the analysis yielding the lowest expected \cls (among those considered) excludes it. The impact of additional constraints from electroweak precision measurements, flavour observables and dark-matter direct detection experiments was also considered.
The technical aspects of processing the large number of models are a testament to the evolution of the reinterpretation and recasting tools that the ATLAS Collaboration produced during \RunTwo~\cite{ATL-PHYS-PUB-2022-017,ATL-PHYS-PUB-2021-038,,ATL-PHYS-PUB-2019-029}, but their discussion is beyond the scope of this paper.

Figure~\ref{fig:pmssm_frac_EWKino} shows, as a function of the electroweakino mass, the fraction of models in the EWKino sample that pass all external constraints but are excluded by ATLAS. In general, the excluded fraction decreases with increasing sparticle mass. About 50\% of the models featuring a neutralino LSP with a mass of $500\,\GeV$ are excluded by ATLAS. This fraction increases to about 80\% for masses of $200\,\GeV$. Heavier electroweakinos with large masses are excluded in a significant fraction of the models: this is often the effect of correlations with the lighter states. For example, in models with a long-lived wino-like lightest chargino, its lifetime is determined by the mass of $\ninotwo$.

\begin{figure}[t]
\centering
\includegraphics[width=0.6\textwidth]{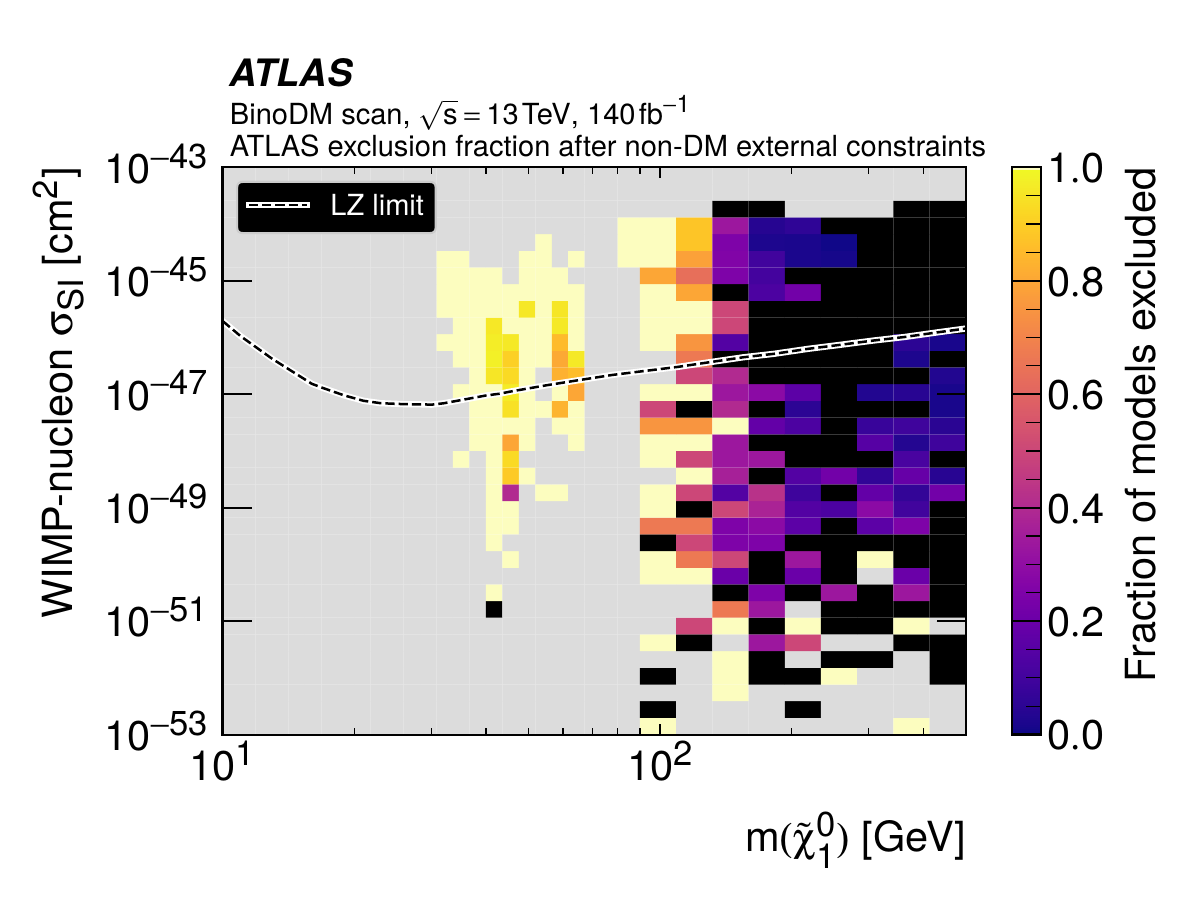}
\caption{Fraction of models passing all non-DM external constraints but excluded by ATLAS. The $y$-axis shows  the WIMP--nucleon spin-independent scattering cross-section $\sigma_{\mathrm{SI}}$, while the $x$-axis shows the mass of the neutralino LSP $\mass{\ninoone}$. Bins in grey indicate that no model passes external constraints, while for bins in cream (black) all models are (not) excluded.}
\label{fig:pmssm_dm}
\end{figure}

A complementary view is given in Figure~\ref{fig:pmssm_dm}: here the Bino-DM sample is considered, and the fraction of models excluded is given in the format usually used by experiments doing direct searches for dark matter, i.e.\ in the plane of the Weakly Interacting Massive Particle (WIMP)--nucleon spin-independent scattering cross-section versus the mass of the WIMP. ATLAS excludes a very large fraction of the $Z/h$ funnel, where the WIMP mass is close to half the mass of one of the bosons, so that the WIMP self-interaction is regulated by an enhanced cross-section for  annihilation via boson exchange. A large fraction of the models are generally excluded for WIMP masses up to about $150\,\GeV$, while the exclusion is progressively more limited above this WIMP mass.

A very useful by-product of studies like the one in Ref.~\cite{SUSY-2020-15} is the identification of classes of models that are not excluded by ATLAS, despite the masses of the involved sparticles being well within the exclusion limits provided by the simplified models. This information was used at the end of \RunOne to design some of the simplified models for \RunTwo, and will be used again to guide the evolution of the research programme in \RunThr and beyond.


\FloatBarrier
\section{Discussion and conclusions}
%

%

The main results of searches for direct production of supersymmetric particles by the ATLAS Collaboration using the \RunTwo dataset are summarised in this paper. The authors hope that the material will provide a useful overview for the casual reader, and a handy collection of references for those actively interested in the subject.

The achievements of the ATLAS Collaboration (and of its counterpart, CMS) are remarkable, and exclude important regions of the parameter space. The question that anyone reading this paper is likely to have in mind at this point is: where does this experimental effort leave the community in terms of the existence of SUSY in nature?

In this form, the question is simply too wide to be answered in a convincing way. Since it is probably impossible to prove that the mass of the photon is exactly zero, it is experimentally impossible to prove that the superpartners of the SM particles do not exist, not even at some very large mass scale. However, the community has concluded that the mass of the photon is small enough to not have any phenomenological consequences for a large variety of phenomena at a cosmological scale. Likewise, one can try to provide answers to specific questions once they are asked in the framework of a supersymmetric theory.

The pre-LHC SUSY landscape was dominated by a relatively small number of frameworks arising from specific top-down approaches to the way SUSY was broken. Models that belong to this family are the constrained MSSM (cMSSM, or mSUGRA), anomaly-mediated SUSY breaking (AMSB), and gauge-mediated SUSY breaking (GMSB). In their minimal forms, the predictions of these models were already under very severe pressure by the end of \RunOne. The limits on the masses of the gluinos and squarks (already in the $\TeV$ regime at the time) excluded a vast amount of the parameter space for the cMSSM~\cite{Buchmueller:2013rsa}. The limits on pure-wino dark matter imposed by the disappearing-track results severely constrained the existence of long-lived charginos, and the difficulty in accommodating a Higgs boson with a mass of $125\,\GeV$ strongly excluded AMSB as a viable framework~\cite{Baer:2018hwa}. Similarly, minimal GMSB needs top squarks with very high mass to obtain a Higgs boson mass compatible with that observed~\cite{Delgado:2015bwa}. Beyond the need to connect searches to specific topologies, the rise of simplified models in \RunTwo is also partly due to the heavy pressure of the \RunOne searches and Higgs boson results on mainstream SUSY-breaking models.   In short, the \RunOne results changed the theoretical thinking and opened up the parameter space.

The \RunTwo research programme was driven by a more agnostic approach -- the big questions (such as: Is SUSY the way nature has chosen to stabilise the Higgs boson mass at the electroweak scale? Is SUSY the reason for the existence of dark matter?) were still used as a guideline, but the approach was much closer to \enquote{turn every stone} than in the past. The classical paradigm of SUSY as a solution to the little hierarchy problem\footnote{SUSY removes the quadratic sensitivity of the weak scale to the presence of a significantly higher energy scale coupling to it (big hierarchy). The residual tuning required between the parameters of a SUSY model to yield the observed weak scale is referred to as the little hierarchy problem.}~\cite{Barbieri:1987fn,deCarlos:1993yy} requires higgsinos with masses of at most a few hundred $\GeV$, top squarks at the $\TeV$ scale, and gluinos not too far above that. The current limits, well into the few-$\TeV$ region for gluinos and a $\TeV$ for top squarks, exceed the expectations for classical naturalness definitions, although more modern reanalysis of the arguments relaxes the constraints~\cite{Baer:2017yqq}. For the first time, hadron collider experiments started to surpass the LEP limits on higgsinos -- and further investigations into the existence of higgsinos will be a highlight of \RunThr and beyond. As discussed in Section~\ref{sec:pmssm}, many scenarios that include viable dark-matter candidates have been severely constrained by the LHC results, most notably the $Z$/$h$ funnels for bino dark matter (those models where the self-annihilation cross-section for binos is enhanced by the proximity of the neutralino mass to half that of the $Z$ or $h$ boson). First sensitivity to the existence of a light stau was achieved during \RunTwo, although the interesting region for probing the stau funnel as a dark-matter regulation mechanism is still beyond experimental reach.

These results have led to a culture shift in the experimental and phenomenology communities: from being the prime candidate as a framework for physics beyond the SM, SUSY at the electroweak scale seems now to be perceived as a somewhat disfavoured candidate mechanism to extend the SM. Is this shift fully justified? The question is subject of an active debate in the theory community (scenarios arising from the string landscape agree well with the current LHC null result~\cite{Baer:2020kwz}).  Regardless of the community's perception, the MSSM scans presented here and independently performed by the phenomenology community clearly show that SUSY still has a lot to offer, both as a source of viable models of new physics and as a spectacular tool for generating new signatures, and is thus a source of new ideas for searches. Hopefully, one of the benefits of reviews like this one is to help the average high-energy physicist to put the experimental effort in the right context: despite its main incarnations having been severely impacted by data from the LHC, the framework of electroweak-scale SUSY still offers viable extensions of the SM.

Together with those from the CMS Collaboration, the results summarised in this paper represent the state of the art (at the time of writing) in the worldwide experimental search for SUSY. No large increase in the centre-of-mass energy of the collisions is foreseen (in the ongoing \RunThr, protons are collided at  $\sqrt{s} = 13.6\,\TeV$). While it will take a long time for the size of the available dataset to increase substantially, the authors are confident that the creativity and skill of the experimentalists will improve the sensitivity of ATLAS to SUSY particles at a rate well beyond simple integrated luminosity scaling, as has happened so many times in the past. Therefore, while the results discussed in this paper are based on much of the overall capability of the LHC to probe unexplored territory in the SUSY parameter space, experimental breakthroughs will open new sensitivity windows, and, hopefully, start to shed some light on the questions that still today, more than ever, call for an extension of the Standard Model of particle physics.


\section*{Acknowledgements}

%
%

%
%

We thank CERN for the very successful operation of the LHC and its injectors, as well as the support staff at
CERN and at our institutions worldwide without whom ATLAS could not be operated efficiently.

The crucial computing support from all WLCG partners is acknowledged gratefully, in particular from CERN, the ATLAS Tier-1 facilities at TRIUMF/SFU (Canada), NDGF (Denmark, Norway, Sweden), CC-IN2P3 (France), KIT/GridKA (Germany), INFN-CNAF (Italy), NL-T1 (Netherlands), PIC (Spain), RAL (UK) and BNL (USA), the Tier-2 facilities worldwide and large non-WLCG resource providers. Major contributors of computing resources are listed in Ref.~\cite{ATL-SOFT-PUB-2023-001}.

We gratefully acknowledge the support of ANPCyT, Argentina; YerPhI, Armenia; ARC, Australia; BMWFW and FWF, Austria; ANAS, Azerbaijan; CNPq and FAPESP, Brazil; NSERC, NRC and CFI, Canada; CERN; ANID, Chile; CAS, MOST and NSFC, China; Minciencias, Colombia; MEYS CR, Czech Republic; DNRF and DNSRC, Denmark; IN2P3-CNRS and CEA-DRF/IRFU, France; SRNSFG, Georgia; BMBF, HGF and MPG, Germany; GSRI, Greece; RGC and Hong Kong SAR, China; ISF and Benoziyo Center, Israel; INFN, Italy; MEXT and JSPS, Japan; CNRST, Morocco; NWO, Netherlands; RCN, Norway; MEiN, Poland; FCT, Portugal; MNE/IFA, Romania; MESTD, Serbia; MSSR, Slovakia; ARIS and MVZI, Slovenia; DSI/NRF, South Africa; MICIU/AEI, Spain; SRC and Wallenberg Foundation, Sweden; SERI, SNSF and Cantons of Bern and Geneva, Switzerland; NSTC, Taipei; TENMAK, T\"urkiye; STFC/UKRI, United Kingdom; DOE and NSF, United States of America.

Individual groups and members have received support from BCKDF, CANARIE, CRC and DRAC, Canada; CERN-CZ, PRIMUS 21/SCI/017 and UNCE SCI/013, Czech Republic; COST, ERC, ERDF, Horizon 2020, ICSC-NextGenerationEU and Marie Sk{\l}odowska-Curie Actions, European Union; Investissements d'Avenir Labex, Investissements d'Avenir Idex and ANR, France; DFG and AvH Foundation, Germany; Herakleitos, Thales and Aristeia programmes co-financed by EU-ESF and the Greek NSRF, Greece; BSF-NSF and MINERVA, Israel; Norwegian Financial Mechanism 2014-2021, Norway; NCN and NAWA, Poland; La Caixa Banking Foundation, CERCA Programme Generalitat de Catalunya and PROMETEO and GenT Programmes Generalitat Valenciana, Spain; G\"{o}ran Gustafssons Stiftelse, Sweden; The Royal Society and Leverhulme Trust, United Kingdom.

In addition, individual members wish to acknowledge support from CERN: European Organization for Nuclear Research (CERN PJAS); Chile: Agencia Nacional de Investigaci\'on y Desarrollo (FONDECYT 1190886, FONDECYT 1210400, FONDECYT 1230812, FONDECYT 1230987); China: National Natural Science Foundation of China (NSFC - 12175119, NSFC 12275265, NSFC-12075060); Czech Republic: PRIMUS Research Programme (PRIMUS/21/SCI/017); EU: H2020 European Research Council (ERC - 101002463); European Union: European Research Council (ERC - 948254, ERC 101089007), Horizon 2020 Framework Programme (MUCCA - CHIST-ERA-19-XAI-00), European Union, Future Artificial Intelligence Research (FAIR-NextGenerationEU PE00000013), Italian Center for High Performance Computing, Big Data and Quantum Computing (ICSC, NextGenerationEU); France: Agence Nationale de la Recherche (ANR-20-CE31-0013, ANR-21-CE31-0013, ANR-21-CE31-0022, ANR-22-EDIR-0002), Investissements d'Avenir Labex (ANR-11-LABX-0012); Germany: Baden-Württemberg Stiftung (BW Stiftung-Postdoc Eliteprogramme), Deutsche Forschungsgemeinschaft (DFG - 469666862, DFG - CR 312/5-2); Italy: Istituto Nazionale di Fisica Nucleare (ICSC, NextGenerationEU), Ministero dell'Università e della Ricerca (PRIN - 20223N7F8K - PNRR M4.C2.1.1); Japan: Japan Society for the Promotion of Science (JSPS KAKENHI JP21H05085, JSPS KAKENHI JP22H01227, JSPS KAKENHI JP22H04944, JSPS KAKENHI JP22KK0227); Netherlands: Netherlands Organisation for Scientific Research (NWO Veni 2020 - VI.Veni.202.179); Norway: Research Council of Norway (RCN-314472); Poland: Polish National Agency for Academic Exchange (PPN/PPO/2020/1/00002/U/00001), Polish National Science Centre (NCN 2021/42/E/ST2/00350, NCN OPUS nr 2022/47/B/ST2/03059, NCN UMO-2019/34/E/ST2/00393, UMO-2020/37/B/ST2/01043, UMO-2021/40/C/ST2/00187, UMO-2022/47/O/ST2/00148); Slovenia: Slovenian Research Agency (ARIS grant J1-3010); Spain: BBVA Foundation (LEO22-1-603), Generalitat Valenciana (Artemisa, FEDER, IDIFEDER/2018/048), La Caixa Banking Foundation (LCF/BQ/PI20/11760025), Ministry of Science and Innovation (MCIN \& NextGenEU PCI2022-135018-2, MICIN \& FEDER PID2021-125273NB, RYC2019-028510-I, RYC2020-030254-I, RYC2021-031273-I, RYC2022-038164-I), PROMETEO and GenT Programmes Generalitat Valenciana (CIDEGENT/2019/023, CIDEGENT/2019/027); Sweden: Swedish Research Council (VR 2018-00482, VR 2022-03845, VR 2022-04683, VR grant 2021-03651), Knut and Alice Wallenberg Foundation (KAW 2017.0100, KAW 2018.0157, KAW 2018.0458, KAW 2019.0447, KAW 2022.0358); Switzerland: Swiss National Science Foundation (SNSF - PCEFP2\_194658); United Kingdom: Leverhulme Trust (Leverhulme Trust RPG-2020-004); United States of America: U.S. Department of Energy (ECA DE-AC02-76SF00515), Neubauer Family Foundation.

%
%


%
%
%
\printbibliography
\clearpage

\clearpage
 
\begin{flushleft}
\hypersetup{urlcolor=black}
{\Large The ATLAS Collaboration}

\bigskip

\AtlasOrcid[0000-0002-6665-4934]{G.~Aad}$^\textrm{\scriptsize 103}$,
\AtlasOrcid[0000-0001-7616-1554]{E.~Aakvaag}$^\textrm{\scriptsize 16}$,
\AtlasOrcid[0000-0002-5888-2734]{B.~Abbott}$^\textrm{\scriptsize 121}$,
\AtlasOrcid[0000-0002-1002-1652]{K.~Abeling}$^\textrm{\scriptsize 55}$,
\AtlasOrcid[0000-0001-5763-2760]{N.J.~Abicht}$^\textrm{\scriptsize 49}$,
\AtlasOrcid[0000-0002-8496-9294]{S.H.~Abidi}$^\textrm{\scriptsize 29}$,
\AtlasOrcid[0009-0003-6578-220X]{M.~Aboelela}$^\textrm{\scriptsize 44}$,
\AtlasOrcid[0000-0002-9987-2292]{A.~Aboulhorma}$^\textrm{\scriptsize 35e}$,
\AtlasOrcid[0000-0001-5329-6640]{H.~Abramowicz}$^\textrm{\scriptsize 152}$,
\AtlasOrcid[0000-0002-1599-2896]{H.~Abreu}$^\textrm{\scriptsize 151}$
% ...
\bigskip
\\

% ...
$^{16}$Department for Physics and Technology, University of Bergen, Bergen; Norway.\\
% ...
$^{29}$Physics Department, Brookhaven National Laboratory, Upton NY; United States of America.\\
% ...
$^{35}$$^{(a)}$Facult\'e des Sciences Ain Chock, Universit\'e Hassan II de Casablanca;$^{(b)}$Facult\'{e} des Sciences, Universit\'{e} Ibn-Tofail, K\'{e}nitra;$^{(c)}$Facult\'e des Sciences Semlalia, Universit\'e Cadi Ayyad, LPHEA-Marrakech;$^{(d)}$LPMR, Facult\'e des Sciences, Universit\'e Mohamed Premier, Oujda;$^{(e)}$Facult\'e des sciences, Universit\'e Mohammed V, Rabat;$^{(f)}$Institute of Applied Physics, Mohammed VI Polytechnic University, Ben Guerir; Morocco.\\
% ...
$^{44}$Physics Department, Southern Methodist University, Dallas TX; United States of America.\\
% ...
$^{49}$Fakult\"{a}t Physik , Technische Universit{\"a}t Dortmund, Dortmund; Germany.\\
% ...
$^{55}$II. Physikalisches Institut, Georg-August-Universit\"{a}t G\"ottingen, G\"ottingen; Germany.\\
% ...
$^{103}$CPPM, Aix-Marseille Universit\'e, CNRS/IN2P3, Marseille; France.\\
% ...
$^{121}$Homer L. Dodge Department of Physics and Astronomy, University of Oklahoma, Norman OK; United States of America.\\
% ...
$^{151}$Department of Physics, Technion, Israel Institute of Technology, Haifa; Israel.\\
$^{152}$Raymond and Beverly Sackler School of Physics and Astronomy, Tel Aviv University, Tel Aviv; Israel.\\
% ...
$^{a}$ Also Affiliated with an institute covered by a cooperation agreement with CERN.\\
$^{b}$ Also at An-Najah National University, Nablus; Palestine.\\
$^{c}$ Also at Borough of Manhattan Community College, City University of New York, New York NY; United States of America.\\
$^{d}$ Also at Center for High Energy Physics, Peking University; China.\\
$^{e}$ Also at Center for Interdisciplinary Research and Innovation (CIRI-AUTH), Thessaloniki; Greece.\\
$^{f}$ Also at Centro Studi e Ricerche Enrico Fermi; Italy.\\
% ...

\end{flushleft}


\end{document}